\documentclass[final,5p,times,twocolumn]{elsarticle}

\usepackage{lineno,hyperref, url}
\modulolinenumbers[5]
\usepackage{float}
\usepackage{graphicx}
\usepackage{dcolumn}
\usepackage{bm}
\usepackage{amsmath}
\usepackage{amssymb}
\usepackage{booktabs}
\usepackage{tabularx}
\usepackage{rotating}
\DeclareMathOperator\erfc{erfc}
\journal{Journal of High Energy Astrophysics}

\begin{document}

\begin{frontmatter}

\title{Positron Excess from Cosmic Ray Interactions in Galactic Molecular Clouds}

\author[1]{Agnibha De Sarkar}
\ead{agnibha@rri.res.in}
\author[1]{Sayan Biswas}
\ead{sayan@rri.res.in}
\author[1]{Nayantara Gupta}
\ead{nayan@rri.res.in}
\address[1]{Raman Research Institute, C. V. Raman Avenue, Sadashivanagar, Bangalore, 560080, India}

\begin{abstract}
The recent data on cosmic ray positron flux measured near the Earth by the \textit{Alpha Magnetic Spectrometer} (AMS-02) experiment extends to TeV energy. {The recently observed data by AMS-02 clearly confirms that the positron flux rises with energy and shows a peak near a few hundred GeV}. This rising positron flux cannot be explained  by  interactions of cosmic rays with interstellar hydrogen gas. {Pulsars, dark matter and many other innovative physical scenarios have been studied to explain this rising of the positron flux, also known as positron excess}. {In this paper, our goal is to study whether secondary production due to cosmic ray interactions in nearby Galactic Molecular Clouds (GMCs) can contribute {significantly} to the observed positron spectrum on Earth.} Due to the progress in multi-wavelength astronomy, many new GMCs have been discovered in our Galaxy recently. {Using large scale CO survey, 1064 GMCs were detected in the Galaxy, which reside in the Galactic plane. Alongside that, very recent survey implemented the optical/IR dust extinction measurements, to trace 567 GMCs within 4 kpc of Earth, also residing in the Galactic plane.} We use the updated list of GMCs {reported in recent papers}, which are distributed in the Galactic plane, to find the secondary positrons produced in them in interactions of cosmic rays with molecular hydrogen.  Moreover, by analysing the \textit{Fermi}-LAT data, new GMCs have been discovered near the Galactic plane. We also include some of these GMCs closest to the Earth where cosmic ray interactions are producing secondaries. It has been speculated earlier that cosmic rays may be reaccelerated in some GMCs. {We select 7 GMCs out of 567 GMCs recently reported, within 4 kpc of Earth, where reacceleration due to magnetized turbulence is assumed}. {We include a small hardened component of secondary positrons, produced from interaction of reaccelerated CRs in those 7  GMCs}. {We use publicly available code \textbf{DRAGON} for our simulation setup to study CR propagation in the Galaxy and show that the observed positron spectrum can be well explained in the energy range of 1 to 1000 GeV by our self consistent model.}
\end{abstract}


\end{frontmatter}

\section{\label{sec:intro}Introduction}
Galactic cosmic rays (CRs) are generally considered to be accelerated in shocks near the supernova remnants (SNRs) \cite{bell78a,bell78b,blandford87,ginzburg64,berezinskii90} and propagated throughout the Galaxy. During their propagation, they are deflected by the Galactic magnetic field (GMF), and they also interact with interstellar hydrogen gas. The secondary CRs, produced in subsequent interactions of primary CRs with interstellar hydrogen gas \cite{strong07}, are important probes of CR acceleration. Other than the interaction with distribution of interstellar matter, there is also diffusion of CRs in the Galaxy, which depends on the magnetic field structure.  
Even after more than a hundred year of discovery of CRs, new observational data brings in new challenges for theoretical interpretations \cite{blasi13}, due to this reason, this field has remained an active area of research.

Electrons are injected by CR sources, also they are produced in interactions of CR protons and nuclei with interstellar matter during their propagation in the Galaxy. While CR protons and nuclei can propagate long distances without losing energy significantly, electrons lose energy within a much shorter distance due to radiative losses. Positrons and antiprotons are secondary particles produced in interactions of CR protons and nuclei with interstellar matter. Being antiparticles, they are useful probes of new physics. 

\textit{Payload for Antimatter Matter Exploration and Light-nuclei Astrophysics} (PAMELA) is a satellite-borne apparatus  for recording charged CRs. The  positron fraction measured by PAMELA  between 1.5 and 100 GeV was the first result showing deviation from the conventional secondary production model \cite{adriani09, adriani10}. The more recent results of PAMELA confirm that additional sources, either astrophysical or exotic, may be required to explain the CR positron spectra \cite{adriani13}.

\textit{Fermi}-LAT collaboration reported the CR  electron and positron spectrum separately and also the positron fraction in the energy range of 20-200 GeV \cite{ackermann12}. They confirmed that the positron fraction rises with the energy in 20-100 GeV energy range, and the three spectral points in that spectrum between 100 and 200 GeV are also consistent with the same feature.

The \textit{Alpha Magnetic Spectrometer} (AMS) on the International Space Station has measured CR fluxes with high precision over a wide energy range. The AMS collaboration published their results on high precision measurements of fluxes of CR protons (p) \cite{aguilar15a}, helium (He)\cite{aguilar15b}, Boron (B) to Carbon (C) flux ratio \cite{aguilar16a}, and also antiprotons ($\rm{\bar{p}}$) \cite{aguilar16b}. Their first results on precise measurements of positron fraction in primary CRs in the energy range of 0.5-350 GeV showed that the positron fraction steadily increases in the energy range of 10 to 250 GeV, however beyond 20 GeV, the slope decreases by an order of magnitude \cite{aguilar13}. Their subsequent results gave better statistics over an extended energy range \cite{accardo14,aguilar14a,aguilar14b}. Their recent results of CR electron \cite{aguilar19a} and positron spectra \cite{aguilar19b} provide high quality measurements of fluxes up to TeV energy. 
The positron flux plotted in GeV$^2$ m$^{-2}$ sec$^{-1}$ sr$^{-1}$ shows significant excess  starting from 25.2$\pm$1.8 GeV and a sharp decrease above 284$^{+91}_{-64}$ GeV. 
The flux has a cutoff at 810$^{+310}_{-180}$ GeV.
The data shows that at high energy the positrons may be originated either from  dark matter (DM) annihilation or from other astrophysical sources.

The DM origin of positron excess was studied in many earlier papers \cite{cholis09,bergstrom08,porter11,cholis13,malyshev09,lin15}. 
Both DM and pulsar scenarios could be the possible origin of the positron excess. Along with anisotropy, which could be another useful probe to discriminate these two scenarios \cite{grasso09}.
Geminga pulsar has long been identified as a nearby gamma ray source. 
The possibility of explaining the GeV positron excess with the TeV gamma ray source Geminga was explored by Y\"uksel  \textit{et al}. \cite{yuksel09}.  
Hooper \textit{\textit{et al}.} \cite{hooper09} suggested that a significant contribution to the positron flux between 10 to 100 GeV might be originated from mature pulsars such as Geminga and B0656+14. 
\textit{The  Advanced Thin Ionization  Calorimeter} (ATIC) reported a ``bump" in the high energy flux  of electrons and positrons \cite{atic08}. Several candidate pulsars were listed in \cite{profumo12} that could individually  or coherently  contribute to explain the PAMELA and ATIC data.
 After more precise observation by AMS-02, the role of nearby pulsars was further explored and  they were identified as possible origin of the positron excess \cite{venter15,feng16,joshi17}.

Previously, very high energy gamma ray data from \textit{High Altitude Water Cherenkov} (HAWC) \cite{abeysekara17a} detector also indicated that significant high energy positron flux from nearby pulsars such as Monogem and Geminga can explain the positron excess at 10-100 GeV energy range \cite{hooper17}.
However, the recent measurement of surface brightness profile of TeV nebulae surrounding Geminga  and PSR B0656+14 by HAWC \cite{abeysekara17b} suggests inefficient diffusion of particles from these sources.  When the HAWC and \textit{Fermi}-LAT data are combined, Geminga and PSR B0656+14 are disfavoured as major sources of positron excess in the energy range of 50-500 GeV \cite{xi19} for Kolmogorov type diffusion. In a more recent work, the pulsar PSR B1055-52  is found to be a promising source for explaining  positron excess \cite{kunfang19}. In future, gamma ray astronomy can shed more light on the origin of positron excess.

 Micro-quasars were also considered to be viable sources for explaining the positron excess. It was shown that photo-hadronic interactions in the jets of micro-quasars can produce the excess positron flux which can explain the rise above 10 GeV \cite{gupta14}.   
 
{Galactic molecular clouds (GMCs) are dense reservoirs of cold protons, distributed throughout the Galactic plane. Such concentrated clumps of protons can be ideal laboratory for different particle interactions. In this work, we try to construct a self consistent model, in which we show that the secondary positrons produced from interactions of CRs in nearby GMCs can explain the rise of positron flux above 10 GeV. Our self-consistent model of CR propagation also fits the data of  CR electrons, positron fraction, protons, antiprotons, B/C and $^{10}$Be/$^9$Be ratio as measured by AMS-02 and PAMELA.} 

We consider Galactic SNRs as the primary sources of CRs. During their random movement in the interstellar medium (ISM), CRs interact with ambient gas and also inside the GMCs. In this paper, we will represent our analysis by dividing it into three parts, namely CASE 1, CASE 2 and CASE 3. CASE 1 considers  interactions of primary CRs with interstellar hydrogen gas. CASE 2, then, takes into account the interactions inside GMCs residing on the Galactic plane and listed by Rice \textit{\textit{et al}.} \cite{rice16}, {Chen \textit{\textit{et al}.} \cite{chen20} and Aharonian \textit{\textit{et al}.} \cite{aharonian16}. Nearby GMCs in the Gould Belt complex, Taurus, Lupus and Orion A have not been included in CASE 2. Since these three GMCs are nearby, and have been extensively studied with gamma-ray data from \textit{Fermi}-LAT experiment \cite{aharonian16}, the effect of these GMCs needs separate modelling. Also, in earlier works \cite{dogiel87, dogiel90, dogiel05, coutu99, duvernois01}, it has been discussed that CRs can get reaccelerated inside GMCs due to magnetized turbulence. As a result, the CR spectrum will be hardened. The secondary particles produced from interaction of these CRs will also have a hardened spectrum. Based on the following three conditions, (1)  detection incapability of \textit{Fermi}-LAT (2) radius $\geq$ 10 pc and (3) distance from the Earth $\leq$ 1 kpc, we select 7 GMCs from \cite{chen20}, inside which we assume CRs are reaccelerated. This will be discussed in more details in section \ref{sec:3}. We omit these 7 GMCs from CASE 2 too.} We find that the total flux of positrons from CASE 1 and CASE 2 is not sufficient to explain the  positron excess above 10 GeV.  Subsequently, we incorporate the contributions of secondary CRs from three nearby GMCs {Taurus, Lupus, Orion A, and also 7 selected GMCs from \cite{chen20}.
 The lepton flux from these GMCs have been calculated analytically}, which is our CASE 3. We  show that the total flux from CASE 1, CASE 2 and CASE 3 can explain the positron excess {observed by AMS-02 and PAMELA data. Also our model fits the data of proton, antiproton, electron spectra and also B/C and $^{10}$Be/$^9$Be ratio quite well. The possibility of GMCs being important contributors to the observed CR spectra can facilitate future observations and analysis, that will expand the landscape of cosmic ray theory and experiments to a new direction.} 
 
The outline of this paper is as follows. In section \ref{sec:2}, we discuss the model set up for CR propagation to obtain the results from {CR interactions with ISM and GMCs}. In section \ref{sec:3}, we obtain the secondary flux contributions from CR interactions in nearby GMCs. In section \ref{sec:4}, we show the results of our model. All the data used in plots \footnote{All data are taken from the database \cite{maurin14} unless otherwise specified.} are obtained from AMS-02 and PAMELA. We discuss our findings in section \ref{sec:5} and conclude our work in section \ref{sec:6}.      

\section{Modelling of cosmic ray propagation}
\label{sec:2}
\subsection{Model setup}
\label{subsec:2.1}
The propagation of CRs can be studied, for a given source distribution, density distribution of interstellar medium (ISM), GMF and injection spectrum of primary cosmic rays from their sources, by solving the CR transport equation \cite{ginzburg64,berezinskii90}. In the present work, we study high energy CR propagation in our Galaxy, by solving the transport equation numerically, using publicly available code \textbf{\textbf{DRAGON}}\footnote{The 3D version of the \textbf{\textbf{DRAGON}} code is available at https://github.com/cosmicrays/DRAGON for download.} (Diffusion of cosmic RAys in Galaxy modelizatiON) \cite{bernardo10,evoli17,gaggero12}. \textbf{\textbf{DRAGON}} incorporates various physical processes such as propagation and scattering of CRs in regular and turbulent magnetic fields, CRs interacting with ISM and GMCs, energy losses due to radioactive decay of the nuclei, ionisation loss, Coulomb loss, Bremsstrahlung loss, synchrotron and IC loss, re-acceleration and convection in the Galactic medium, to obtain the solution of the transport equation for the CR propagation in the Galaxy. In this subsection, we give an overview of the source distribution model, GMF model, ISM gas density distribution and diffusion coefficient, that we have chosen for our work.

\textbf{\textbf{DRAGON}} solves the transport equation in 3D geometry, where the Galaxy is assumed to be cylindrical in shape. The outermost radial boundary is denoted as $R_{max}$, vertical boundary as L, and halo height as $z_t$, where $L=3 z_t$ \cite{bernardo13}. The location of the observer is specified at Sun's position with respect to the Galactic center (GC), with x = 8.3 kpc, y = 0, z = 0. We are propagating CRs with atomic number ranging from Z = 1 to Z = 14, considering  propagation of particles with higher mass numbers does not affect our results. Primary CRs in our work, are assumed to be produced from SNRs in our Galaxy. Assuming SNRs as the major sources of CRs with an universal injection spectrum, the source term is used from the paper by K. Ferriere \cite{ferriere01}.

Interstellar gas plays an important role in the process of CR interactions and secondary production. During propagation, CRs interact with different gas components of the ISM. The gaseous components are mainly atomic hydrogen (HI), ionized hydrogen (HII) and molecular hydrogen ($\rm{H_{2}}$). As discussed earlier, we divide the contribution from the interaction of primary CRs with these components into three cases, CASE 1 for contribution from ISM gas density distribution, and CASE 2 for contribution from interactions in GMCs listed in Rice \textit{\textit{et al}.} \cite{rice16},{Chen \textit{\textit{et al}.} \cite{chen20} and Aharonian \textit{\textit{et al}.} \cite{aharonian16}, apart from Taurus, Lupus, Orion A, and 7 GMCs selected from \cite{chen20}. These 10 GMCs are modelled separately as our CASE 3.}  

\textit{\textbf{HI density distribution}}: Neutral or atomic hydrogen cannot be detected in optical wavelengths. Generally, HI can be detected by the observation of Lyman $\alpha$ \cite{savage72,jenkins74} and 21-cm line \cite{dickey90,cox86}.   Previously, many models have been given to describe HI gas distribution \cite{nakanishi03,ferriere98,pohl08}. In our calculation, the radial dependence of HI number density in the Galactic plane is defined by a table in ref. \cite{gordon76}, which is renormalized to make it consistent with the data of ref. \cite{dickey90}.  The z-dependence is calculated using the approximation by \cite{dickey90} for R $<$ 8 kpc, by \cite{cox86} for R $>$ 10 kpc, and interpolated in between.

\textit{\textbf{HII density distribution}}: Radio signals from pulsars and other Galactic and extragalactic compact objects give us the information about the ionized component of hydrogen gas. Some of the models for the distribution of ionized hydrogen component are \cite{ferriere98,ne2001a,ne2001b}. Cordes \textit{\textit{et al}.} \cite{cordes91} provided the space averaged free electron density depending on dispersion, distance and scattering measurements of pulsars. The distribution of ionized component HII is calculated using the cylindrically symmetrical model for space averaged free electron density \cite{cordes91}.

\textit{\textbf{$H_2$ density distribution}}: Molecular hydrogen ($H_2$) is the most abundant molecule in our Galaxy. Second most abundant molecule is CO. Study of $H_2$ cannot be done reliably from UV and optical observations, because UV and optical observations suffer from interstellar extinction. $H_2$ is studied indirectly by radio observation of CO molecules as CO molecule has (J = 1 $\to$ 0) rotational transition at radio wavelength of 2.6 mm \cite{ferriere01}. Such transition of CO acts as a tracer of $H_2$, where CO-to-$\rm{H_{2}}$ conversion factor, $X_{CO}$ is used to obtain information of $H_2$ distribution in Galaxy \cite{ferriere98,pohl08,bronfman88,nakanishi06,ferriere07}. Most of the $H_2$ contribution in our Galaxy comes from GMCs. The radial distribution of GMCs is discussed in subsection \ref{subsec:2.2}. 

Galactic magnetic field (GMF) plays a crucial role in CR propagation. CR leptons lose energy by synchrotron emission in GMF. There are several methods to constrain the intensity and the orientation of GMF: Zeeman splitting observations \cite{crutcher91}, infrared, synchrotron and starlight polarisation studies \cite{nishiyama10,jaffe10,heiles96}, and Faraday rotation measures of the Galactic and extragalactic sources \cite{han06,pshirkov11}. The Galactic magnetic field $\vec{B}$ is usually described as a sum of two components: a large scale regular, and a small scale turbulent, both having a strength of the order of $\mu$G in the Galaxy \cite{beck09}.

In this work, we use the GMF model as given by \cite{pshirkov11}.  The GMF has three components, namely disc, halo and turbulent. The normalizations of the three components are denoted as $B_{0}^{\rm{disc}}$, $B_{0}^{\rm{halo}}$ and $B_{0}^{\rm{turbulent}}$ respectively. $B_{0}^{\rm{disc}}$ and $B_{0}^{\rm{halo}}$ lie in the range of 2-11 $\mu$G but their role in CR propagation is insignificant \cite{bernardo13}.  Among these components, the turbulent component of the GMF plays an important role in CR propagation. Observationally, the most relevant information of the turbulent component of GMF comes from  Faraday Rotation measurements. A functional relation between magnitude of the turbulent magnetic field and halo height ($z_{t}$) is given in \cite{bernardo13} by theoretical modeling of the propagation of the Galactic CR electrons and positrons to fit their observed fluxes, their synchrotron emission and its angular distribution. This expression from \cite{bernardo13},
\begin{equation}
\label{eq1}
\left(\frac{B^{turbulent}_0}{1\:\mu G}\right)^2 = 148.06\:\left(\frac{1\:kpc}{z_t}\right) + 19.12\
\end{equation}
  has been used to calculate the intensity of random magnetic field. The shape of the vertical profile is poorly constrained. We have used exponential profile of the random component of the magnetic field, which is compatible with presently available data. 

 In the present work, we have used the following form of diffusion coefficient to study the CR propagation in the Milky Way Galaxy. 
  
\begin{equation}
\label{eq2}
D(\rho, z) = \beta^\eta D_0 \left(\frac{\rho}{\rho_0}\right)^\delta exp\left(\frac{z}{z_t}\right)\,,
\end{equation}
where, $\rho$ being the rigidity, $z$ is the vertical height above the Galactic plane and $\delta$ denotes the power law index. $z_t$ and $\beta$ are Galactic halo height and dimensionless particle velocity respectively. The power $\eta$ of $\beta$ accounts for the uncertainties that arise due to propagation of CRs at low energies \cite{bernardo11}. $D_0$ denotes the normalisation of diffusion coefficient and $\rho_0$ is the reference rigidity. Also note that, to avoid the boundary effects, we set L = 3$z_t$ in our work \cite{bernardo13}. The z-component of diffusion coefficient and the turbulent magnetic field are related by 

\begin{equation}
\label{eq3}
D(z)^{-1} \propto B^{\rm{turbulent}} (z) \propto exp(-z/z_{t}).
\end{equation} 


In our study, we have used injection spectra of protons and heavy nuclei in the following form  \cite{cholis12}:
\begin{equation}
\label{eq4}
\begin{split}
\frac{dN^{k}}{d\rho} \propto 
\begin{cases}
(\rho/\rho_{br, 1}^{k})^{-\alpha_1^k}& \rho < \rho_{br, 1}^{k}\\
(\rho/\rho_{br, 1}^k)^{-\alpha_2^k} &  \rho_{br, 1}^k \leq\:\rho\\
\end{cases}              
\end{split}
\end{equation}
In \textbf{\textbf{DRAGON}}, $\alpha_1^k, \alpha_2^k, \rho_{br, 1}^k$ are free parameters, which have been tuned to fit the observed CR spectra. In the above relation, k denotes protons and heavy nuclei ($k=1,2,..,14$), whose spectra, we assumed to be similar in our case. Similarly, for electron injection spectra, we use a similar form:
\begin{equation}
\label{eq5}
\begin{split}
\frac{dN^{e}}{d\rho} \propto 
\begin{cases}
(\rho/\rho_{br, 1}^{e})^{-\alpha_1^e}& \rho \le \rho_{br, 1}^{e}\\
(\rho/\rho_{br, 1}^e)^{-\alpha_2^e} &  \rho_{br, 1}^e<\:\rho < \rho_{br, 2}^e\\
(\rho/\rho_{br, 2}^e)^{-\alpha_3^e}(\rho_{br, 2}^e/\rho_{br, 1}^e)^{-\alpha_2^e} & \rho_{br, 2}^e\le\:\rho\\
\end{cases}              
\end{split}
\end{equation}
We also need to take into account the solar modulation effect which is dominant below 10 GeV. {In accordance with the force-field approximation, we have implemented the solar modulation with a potential ($\phi$) such that the observed spectrum can be written as, \cite{usoskin05}, 
 \begin{equation}
\label{eq6}
J_k (T_k, \phi) = J_{LIS, k} (T_k + \Phi)\:\frac{(T_k)\:(T_k + 2T_p)}{(T_k + \Phi)(T_k + \Phi + 2T_p)}
\end{equation}
where $\phi$ is the solar modulation potential, $J_k$ is the differential intensity of the CR nuclei, $T_k$ is the kinetic energy of CR nuclei with charge number Z and mass number A and $\Phi = (Ze/A)\:\phi$. $T_p$ is the proton rest mass energy and $J_{LIS, k}$ is local interstellar spectrum of CR nuclei type k. Similarly, for electron and positron, the equation will take the form,
 \begin{equation}
\label{eq7}
J_e (T_e, \phi) = J_{LIS, e} (T_e + \phi)\:\frac{(T_e)\:(T_e + 2T_q)}{(T_e + \phi)(T_e + \phi + 2T_q)}
\end{equation}
where $\phi$ is the solar modulation potential, $J_e$ is the differential intensity of the electron-positron, $T_e$ is the kinetic energy of electron-positron, $T_q$ is the electron rest mass energy and $J_{LIS, e}$ is local interstellar spectrum of electron-positron.}

\subsection{Distribution of Galactic Molecular clouds}
\label{subsec:2.2}
\begin{figure*}
\centering 
\includegraphics[width=.85\textwidth,origin=c,angle=0]{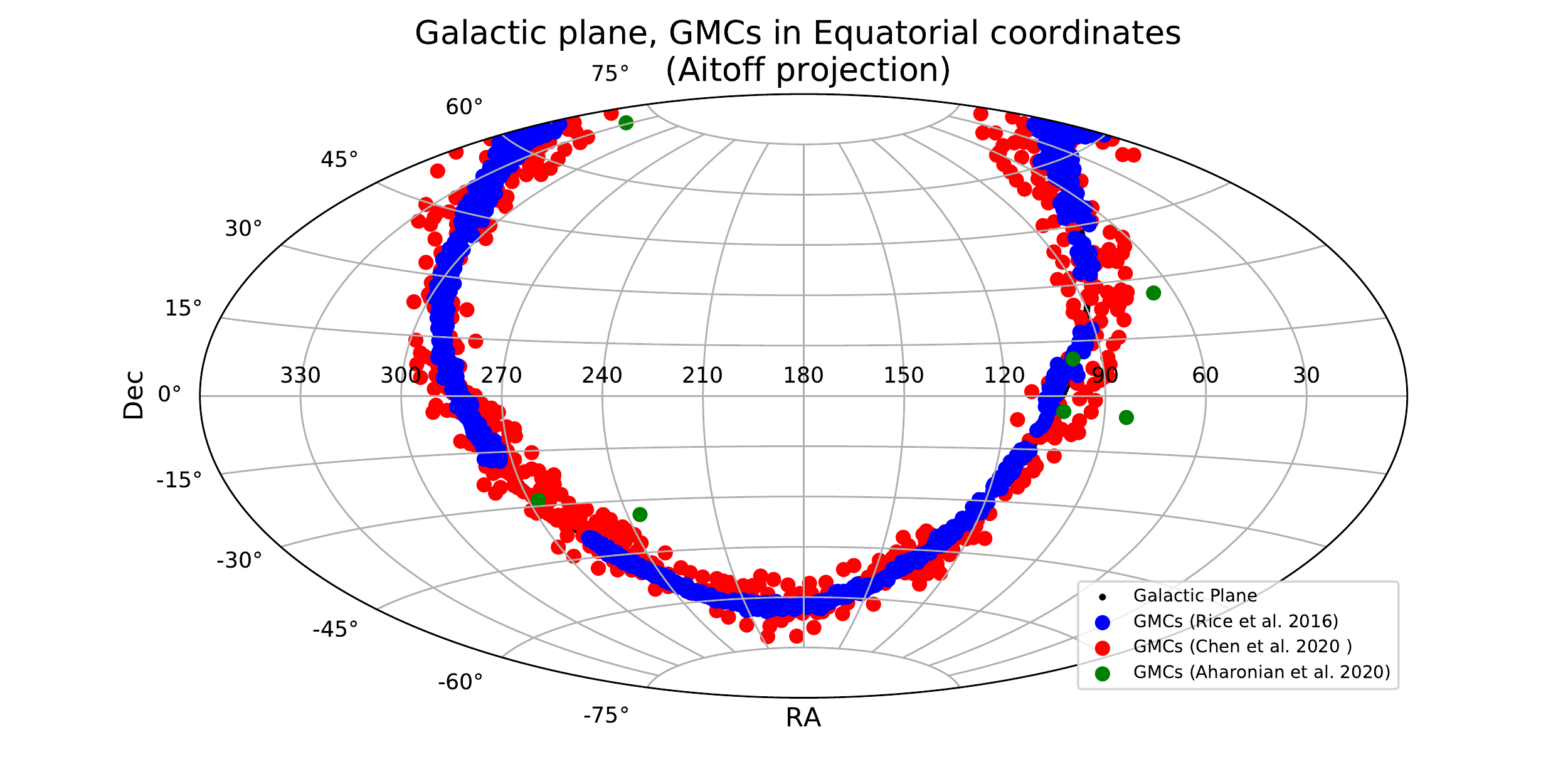}
\caption{\label{fig1} All sky map of the GMCs taken for this work from Rice \textit{et al} \cite{rice16}, Chen \textit{et al} \cite{chen20} and Aharonian \textit{et al} \cite{aharonian16}.}
\end{figure*}
\begin{figure*}
\centering
\includegraphics[width=0.7\textwidth,origin=c,angle=0]{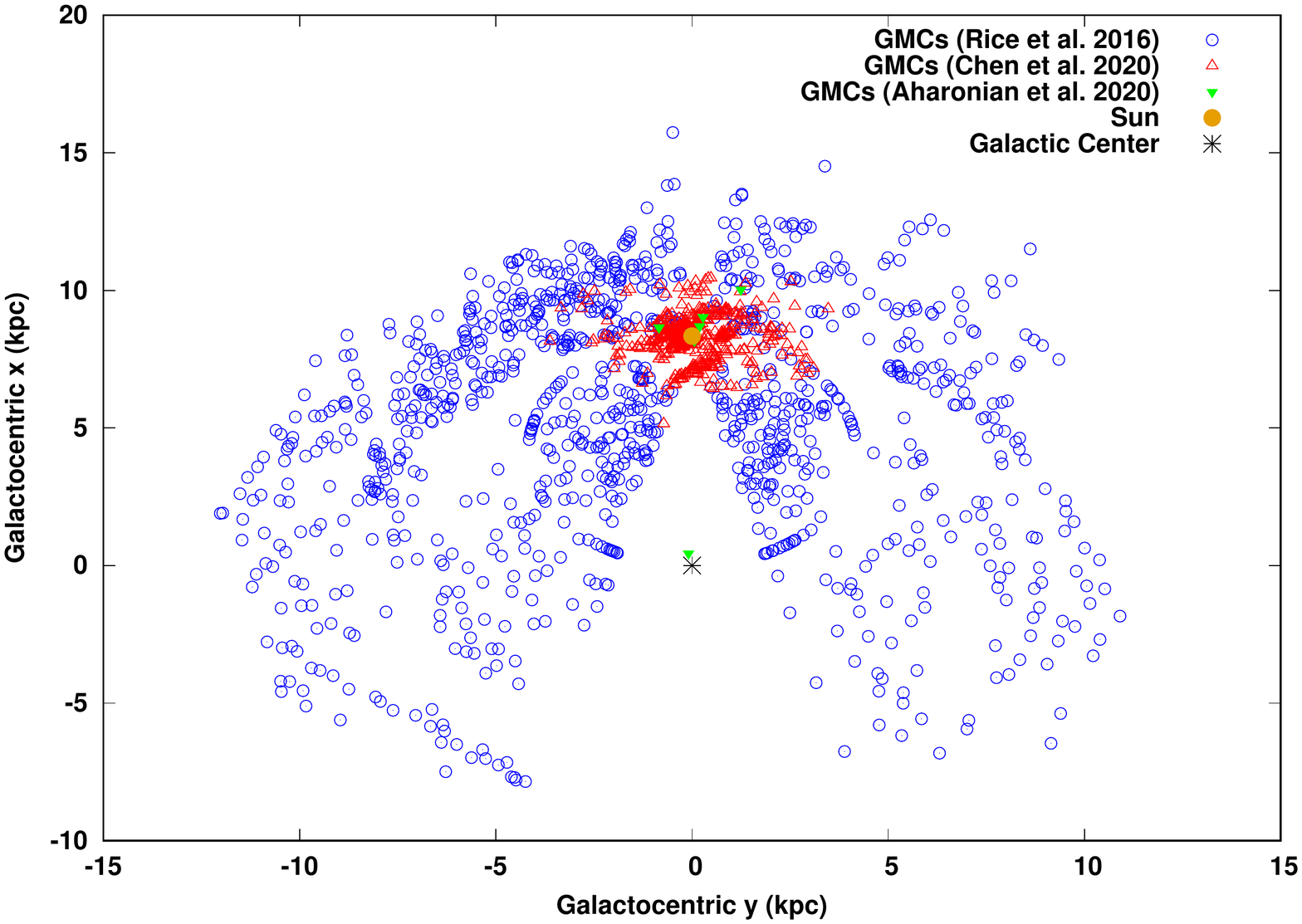}
\caption{\label{fig2} Positional distribution of the GMCs in a 2D X-Y plane.}
\end{figure*}      
\begin{figure}
\includegraphics[width=0.5\textwidth,origin=c,angle=0]{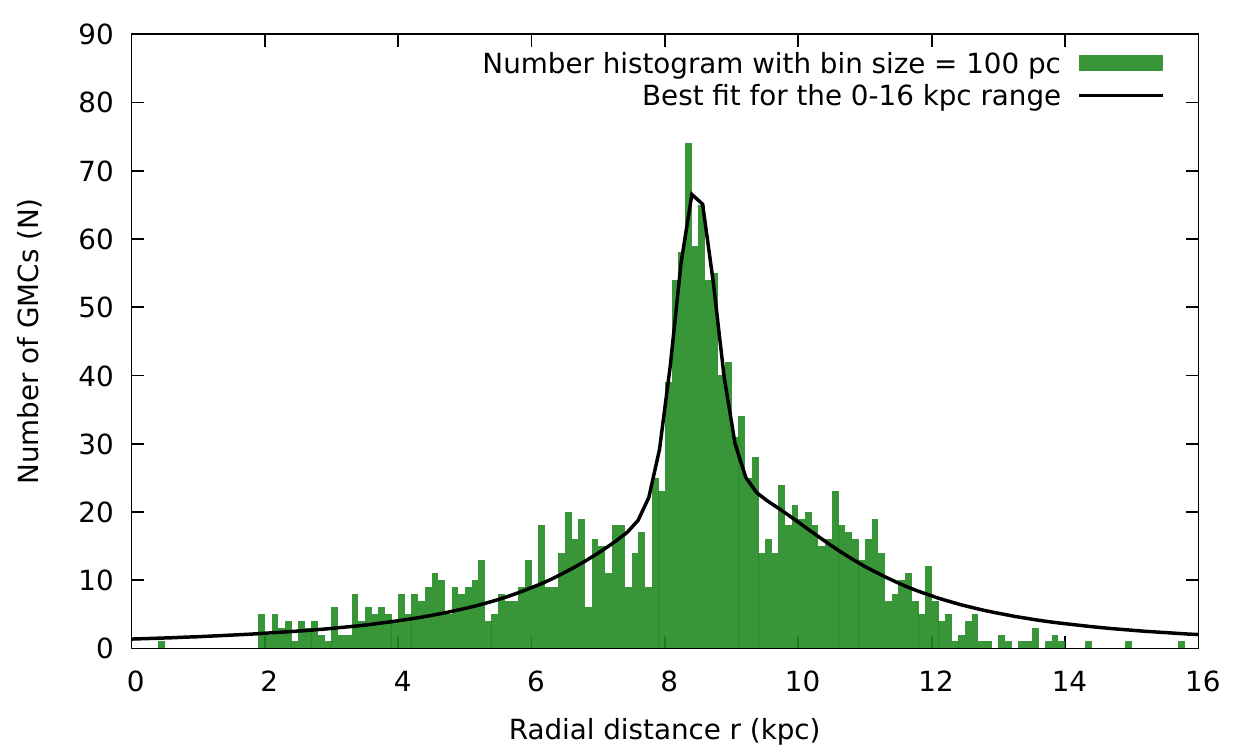}
\caption{\label{fig3} Histogram 1 : Radial number profile of GMCs in Galaxy. All of the GMCs from \cite{rice16}, \cite{chen20} and \cite{aharonian16} have been taken into account in this case. The black line denotes the linear combination of Gaussian and Lorentzian distribution function. This black line depicts the functional fit of the histogram in the entire spatial range.}
\end{figure}
{GMCs are the main sources of molecular hydrogen in our Galaxy. Secondary CRs are produced in interactions of primary CRs in the molecular cloud environment, which contribute to the observed CR spectrum. In this work, we consider all the GMCs recently reported in Rice \textit{\textit{et al}.} \cite{rice16}, Chen \textit{\textit{et al}.} \cite{chen20} and Aharonian \textit{\textit{et al}.} \cite{aharonian16}. First,} we use the catalog of GMCs from Rice \textit{\textit{et al}.} \cite{rice16}, where they presented  a list of 1064 GMCs, by using a dendogram-based decomposition of a previous most uniform, large-scale all-Galaxy CO survey \cite{dame01}. The objects are distributed in the Galactic disk between $180^{\circ} > l > 13^{\circ}$ and $348^{\circ} > l > 180^{\circ}$ within $-5^{\circ} < b < 5^{\circ}$, widely spread in the Galaxy covering distances from $\sim\:$1 to $\sim\:$16 kpc. {Next we take GMCs closer to Earth, which are reported by Chen \textit{\textit{et al}.} \cite{chen20}. These GMCs have been traced by optical/near-infrared (IR) dust extinction measurements. The distances to these GMCs have been accurately measured by 3D dust extinction mapping methods. These GMCs have been identified based on 3D dust reddening maps of the Galactic plane and estimates of colour excess, although these GMCs have not been analyzed by \textit{Fermi}-LAT. In the work by Chen \textit{\textit{et al}.} \cite{chen20}, 567 GMCs have been detected, within 4 kpc from the Earth. The GMCs are distributed in the Galactic disk, in the range of Galactic longitude $0^{\circ} < l <360^{\circ}$ and Galactic latitude $-10^{\circ} < b < 10^{\circ}$. In addition to this, we also take into account the GMCs from the recent work by Aharonian \textit{\textit{et al}.} \cite{aharonian16}, where they have analyzed the \textit{Fermi}-LAT gamma ray data from nearby GMCs.}  Using the information of the Galactic latitude (b) and longitude (l) from these catalogs, we can calculate the positions of the GMCs in the Galaxy in galactocentric coordinate system. 
We use the equations from \cite{rice16,ellsworth13,goodman14}, taking into account that the Sun is at $z_0$ $\sim$ 25 pc above the Galactic plane. The equations are
\begin{equation}
\label{eq8}
\begin{split}
x_{gal} &= R_0\cos{\theta} - d_\odot(\cos{l}\cos{b}\cos{\theta} + \sin{b}\sin{\theta}) \
\\
y_{gal} &= - d_\odot \sin{l} \cos{b} \
\\
z_{gal} &= R_0\sin{\theta} - d_\odot(\cos{l}\cos{b}\sin{\theta} - \sin{b}\cos{\theta}) \
\end{split}
\end{equation}
where, $\theta$ = $sin^{-1}{\frac{z_0}{R_0}}$, $R_0=8.34$ kpc is the distance of the Sun from the GC, $d_\odot$ is the kinematic distance of the individual GMCs from the Sun.
Positional distribution of these GMCs are given in figure~\ref{fig1} and figure~\ref{fig2}.
  
\begin{table*}[htbp]
\caption{\label{tab1} Best fit parameters for linear combination of Gaussian and Lorentzian distributions.}
\centering
\begin{tabular}{ccccccc}
\hline
Histogram & a$_1$ & $\sigma$ & $\mu$ & a$_2$ & r$_0$ & $\gamma$\\
\hline
Histogram 1 & 33.5319 & 0.301458 & 8.45895 & 23.7586 & 8.8329 & 2.19179\\
Histogram 2 & 32.9649 & 0.29976 & 8.45858 & 23.5482 & 8.8312 & 2.20913\\
\hline
\end{tabular}
\end{table*}  
  
Generally, by tracing the CO emission in the Galaxy and multiplying the CO emissivity with the CO-to-$H_{2}$ conversion factor, the gas density of molecular hydrogen is modelled \cite{ferriere98,pohl08,bronfman88,nakanishi06,ferriere07}. {In this work, it can be seen that the GMCs taken from Rice \textit{et al} \cite{rice16}, Chen \textit{et al} \cite{chen20}, and Aharonian \textit{et al} \cite{aharonian16}, predominantly reside in or very close to the Galactic plane}. We consider only radial distribution of the GMCs. We assume concentric circles of constant bin size of 100 pc, centered at the GC, and build  histograms for the number of GMCs residing in each bin in the Galactic plane, covering the radial distance from the GC to the outer region of the Galaxy. The region adjacent to the GC (within $12^{\circ}$) is excluded in the catalog \cite{rice16}, hence there is a large wedge shaped  gap between the first and fourth quadrant, {whereas GMCs reported in Chen \textit{et al} \cite{chen20} span the entire Galactic longitude, hence there is no gap. First, we consider all of the GMCs from Aharonian \textit{et al} \cite{aharonian16}, Rice \textit{et al} \cite{rice16} and Chen \textit{et al} \cite{chen20}, and build number histogram with them, i.e. we consider CASE 2 and CASE 3 GMCs in one histogram. The histogram gives us the variation of the number of GMCs with radial distance. We name this number histogram, where we have included all of the GMCs considered in our work, Histogram 1. Due to inclusion of many GMCs in near Earth region, a peak can be observed near r = 8 kpc. We have fitted the number histogram with a linear combination of Gaussian distribution function and Lorentzian distribution function, which extends from center of the Galaxy to the outer region of the Galaxy. The Gaussian radial distribution function has the general form  $N_1(r) = \frac{a_1}{{\sigma \sqrt {2\pi } }}e^{{{ - \left( {r - \mu } \right)^2 } \mathord{\left/ {\vphantom {{ - \left( {x - \mu } \right)^2 } {2\sigma ^2 }}} \right. \kern-\nulldelimiterspace} {2\sigma ^2 }}}$, where $\sigma$ is the variance, $\mu$ is the mean, `$a_1$' is the normalisation factor, and the Lorentzian radial distribution is given by $N_2 (r) = \frac{a_2\:\gamma^2}{(r - r_0)^2 + \gamma^2}$, where `$a_2$' is the normalisation factor, $r_0$ is  the location of the peak of the distribution, and $\gamma$ denotes half width of the distribution at half of the maximum height. The cumulative distribution function is written by N(r) = N$_1$(r) + N$_2$(r). This distribution function is fitted on the number histogram titled Histogram 1, and the corresponding plot is shown in the figure \ref{fig3}. The fit parameters, for 100 pc bin size, are given in the following table \ref{tab1}. Integrating this number profile N(r) per unit bin width, over the entire spatial region, we get back very closely the total number of GMCs considered. The functional fit of Histogram 1 will be used in determining proton, antiproton fluxes and B/C, $^{10}$Be/$^9$Be ratios.

\begin{figure}
\includegraphics[width=0.5\textwidth,origin=c,angle=0]{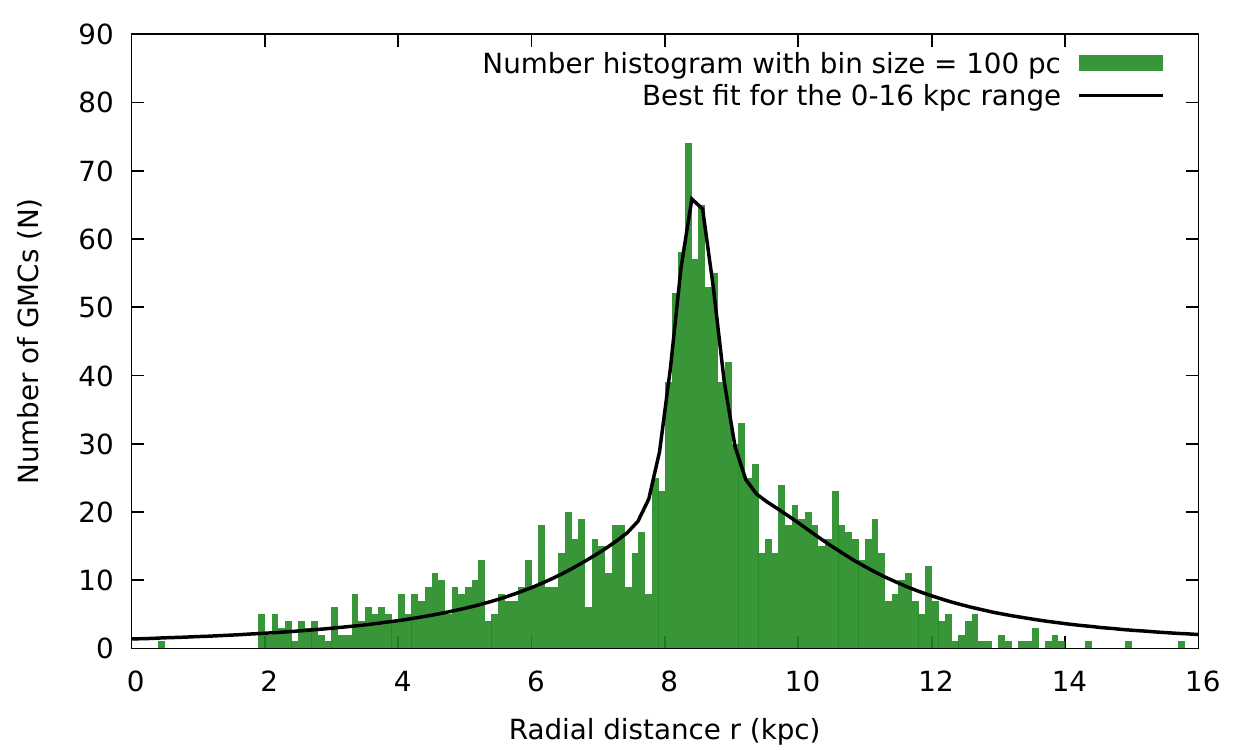}
\caption{\label{fig4} Histogram 2 : Radial number profile of GMCs in Galaxy. All of the GMCs from \cite{rice16}, \cite{chen20} and \cite{aharonian16}, other than 10 GMCs considered for CASE 3, have been taken into account in this case. The black line denotes the linear combination of Gaussian and Lorentzian distribution function. This black line depicts the functional fit of the histogram in the entire spatial range.}
\end{figure}

Next, we consider another scenario, in which we omit nearby GMCs, Taurus, Lupus, Orion A . Moreover, we omit the 7 GMCs out of the 567 GMCs reported in \cite{chen20}, where we have assumed reacceleration due to magnetized turbulence is occuring. These GMCs will be modelled individually in section \ref{sec:3}. Apart from these 10 GMCs, we take all of the other GMCs considered in this work, and following the previous method, build a number histogram, taking 100 pc as bin width, which is our CASE 2. As we can see, the nature of the number histogram in the second scenario does not change much from that of the first scenario, as omission of 10 GMCs does not alter much of the content of the molecular hydrogen component. We label this second number histogram as Histogram 2. As the nature of the Histogram 2 is very similar to that of Histogram 1, we fit this number histogram with the linear combination of Gaussian and Lorentzian distribution function, same as before. The fit parameters are given in table \ref{tab1}, considering 100 pc bin width and the fit of the histogram has been shown in figure \ref{fig4}. As it can be seen, there is a very slight change in the fit parameters, which is expected. This functional fit of the Histogram 2 will be used in determining electron and positron spectra, and positron fraction in the later sections.}

In order to obtain the radial, average $n_{H_2}$ gas density profile in our Galaxy, we have used the following expression,

\begin{equation}
\label{eq9}
\begin{split}
n_{H_2}(r) &= <n_{H_2}> \times \left(\frac{N(r)}{N_{total}}\right) \,,
\end{split}
\end{equation}   

where N(r) represents the {linear combination of Gaussian distribution fits and Lorentzian distribution fits for the number of GMCs in the Galaxy considered in our work, $N_{total}$ is total number of GMCs considered}, and $<n_{H_2}>$ is the average number density. The number density generally considered for GMCs is $\sim$ 100 $cm^{-3}$ \cite{heyer15}. Since we have essentially smoothed out each discrete clumps of GMCs into a radially, continuous distribution of molecular hydrogen, ranging from $\sim$ 0 kpc to $\sim$ 16 kpc, the average density of the distribution is taken as $<n_{H_2}>\:\sim 10\:cm^{-3}$. 

The inclusion of vertical distribution of GMCs does not affect the density distribution used in our study, since all the GMCs considered in this work reside near the Galactic plane as previously stated. {While building Histogram 1, all of the GMCs (CASE 2 + CASE 3) were taken into account. But for the second scenario, 10 GMCs were omitted while building the Histogram 2, hence only CASE 2 GMCs were considered. Secondary lepton production in the 10 nearby GMCs omitted from CASE 2, are modelled individually, which is our CASE 3 and has been discussed in more detail in the next section.}

\section{Contributions from nearby, sub-Kpc GMCs}
\label{sec:3}
After combining CASE 1 and CASE 2 we find that the total positron flux is insufficient to fit the observed data.
Hence, in order to fit the observed flux, we consider the contributions from nearby GMCs (d $\le$ 1 kpc), which is defined as CASE 3 previously.
{The CASE 3 includes Taurus, Lupus and Orion A, which are members of Gould Belt complex, and the 7 GMCs selected from the catalog given by Chen \textit{et al} \cite{chen20}.} GMCs are  dense, concentrated clumps of cold protons in the Galaxy. When primary CR protons injected from the SNRs propagate through these clumps of cold protons, gamma rays and leptons are produced by hadronic interactions (\textit{pp}). {Cosmic ray reacceleration is also a proposed mechanism that can occur inside GMCs due to magnetized turbulence \cite{dogiel87}. We include the contribution of individual nearby GMCs to the total lepton spectra and positron fraction.}

First, we include three of the nearby GMCs Taurus, Lupus and Orion A, which act as local sources of secondary CRs, and contribute to the total fluxes of leptons and gamma rays. We note that these GMCs are not included in the catalog of Rice \textit{et al}. \cite{rice16} {or Chen \textit{et al} \cite{chen20}, hence there has been no double counting in our work}. These three GMCs are members of the Gould Belt complex, and  being our nearest GMCs, they contribute significantly to the positron and electron flux. The gamma ray analysis of these GMCs was done in detail in \cite{aharonian16}. 
 Previously, Taurus and  Orion A were studied in \cite{neronov17,yang14}, while Lupus was studied for the first time in \cite{aharonian16}.
 Following the definition given in \cite{aharonian16}, $B \equiv \frac{M_5}{d_{kpc}^2}$, where $M_5 = \frac{M}{10^5\:M_\odot}$ and $d_{kpc} = \frac{d}{1 kpc}$, M is the mass of the GMCs, d is the distance of these three GMCs from the Earth and $M_{\odot}$ is the solar mass, these three GMCs from the Gould Belt complex has `B' parameter sufficiently higher than 1, which makes them detectable by \textit{Fermi}-LAT. The position coordinates, masses, distances from the Earth and GC, values of the parameter `B' of these three GMCs are given in table \ref{tab2}.
\begin{table*}[htbp]
\centering
\caption{\label{tab2} GMC parameters : Galactic coordinates (l, b), masses M, distances from the Earth (d), Galactocentric distance ($R_{GC}$) and the B parameter from \cite{aharonian16} and references therein. }

\begin{tabular}{ccccccc}
\hline
Cloud  & l & b & Mass & d & $R_{GC}$ & B \\
 & (deg) & (deg) & ($10^5 M_\odot$) & (kpc) & (kpc) &\\
\hline
Taurus & 171.6 & -15.8 & 0.11 & 0.141$\pm$0.007 & 8.4 & 5.6 \\
Lupus & 338.9 & 16.5 & 0.04 & 0.189$\pm$0.009 & 8.2 & 1.0 \\
Orion A & 209.1 & -19.9 & 0.55 & 0.43$\pm$0.02 & 8.4 & 3.0\\
\hline
\end{tabular}

\end{table*}   


The gamma ray fluxes produced in these GMCs in \textit{pp} interactions through the production of neutral pions and their subsequent decay have been calculated in \cite{aharonian16}  and fitted to \textit{Fermi}-LAT data. They have calculated the parent CR proton density spectrum $J_p(E_p)$ for each of these GMCs by fitting the observed gamma ray spectrum, 
\begin{equation}
\label{eq10}
\begin{split}
J_p(E_p) = \rho_{0, CR}\:\left(\frac{E_p}{E_0}\right)^{-\alpha}\\          
 \end{split}
 \end{equation}
where $\rho_{0,CR}$ is the normalisation constant, $E_0=10$ GeV is the reference energy, and $\alpha$ is the spectral index. The values for CR proton density $\rho_{0, CR}$ at 10 GeV and spectral index $\alpha$, for the three GMCs used in our work are given in table \ref{tab3}.
\begin{table}[htbp]
\centering
\caption{\label{tab3} The spectral indices and CR proton densities at 10 GeV derived from the gamma ray and CO data at the location of the GMCs \cite{aharonian16}, errors on the normalisation result from the sum in quadrature of the statistical error deriving from the fit and the $30\%$ uncertainty on the B parameter (see Table III of \cite{aharonian16}).}

\begin{tabular}{p{3cm} p{3cm} p{1.5cm}}
\hline
Cloud  & $\rho_{0, CR}$ & $\alpha$ \\
 & [$10^{-12}\:GeV^{-1}\:cm^{-3}$] &\\
\hline
Taurus & 1.43\:$\pm$\:0.5 & 2.89\:$\pm$\:0.05  \\
Lupus & 1.09\:$\pm$\:0.4 & 2.74\:$\pm$\:0.1\\
Orion A & 1.55\:$\pm$\:0.5 & 2.83\:$\pm$\:0.05\\
\hline
\end{tabular}

\end{table}   

In \textit{pp} interactions charged pions are produced along with neutral pions, which subsequently decay to charged muons. Electrons and positrons are produced from decay of  these charged muons. We have calculated the electron and positron fluxes produced in these three GMCs from \textit{pp} interactions following the formalism given in \cite{kelner06} and using the proton density spectrum given in \cite{aharonian16}.{The cross section for the production of leptons were taken from \cite{kelner06}, which can be written as,}
\begin{equation}
\label{eq11}
\begin{split}
\sigma_{inel} (E_p) & = 34.3 + 1.88\:L + 0.25\:L^2\:\text{mb}, \\
                    & \:\:\:\:\:\:\:\:\:\:\:\:\:\:\:\:\text{for}\:E \geq 100\: \text{GeV}\\ 
                    & = (34.3 + 1.88\:L + 0.25\:L^2)\\
                    & \:\:\:\:\:\:\:\:\:\:\:\:\:\:\:\:\times\left[1-\left(\frac{E_{th}}{E_p}\right)^4\right]^2\text{mb},\\ 
                    & \:\:\:\:\:\:\:\:\:\:\:\:\:\:\:\:\text{for}\:E \leq 100\: \text{GeV}
\end{split}
\end{equation}

{where L = ln($E_p$/1 TeV), and $E_{th}$ = $m_p + 2m_{\pi} + m_{\pi}^2/2m_p = 1.22 \times 10^{-3}$ TeV, which is the threshold energy of production of $\pi$ mesons. \cite{kelner06} obtained this approximation with the fit of the numerical data included in the SIBYLL code.} We note that if we include more GMCs which are further away from the Earth and away from the Galactic plane our result does not change significantly, as the effect of the nearest GMCs is most dominant on the electron and positron fluxes.

GMCs are highly magnetized dense astrophysical objects, with complex inhomogenous structures. Parent CRs injected inside GMCs, traverse through weakly ionized turbulent plasma, which also contains high magnetic field. As a result, particle energies may increase or reacceleration may happen due to fluctuating electromagnetic field inside the GMCs. Although gamma ray data analysis has not shown till now, effects of such reacceleration happening inside GMCs, \cite{dogiel87} argued that it is physically possible. The gravitational energy of the GMCs are very high (ranging in $\sim$ 10$^{50}$ - 10$^{51}$ erg). If at least a very small part of this gravitational energy can be transformed into the energy of the accelerated particles, then the GMCs can effectively act as particle accelerators. It is necessary that the GMCs have to be strongly turbulent in nature, in order to accelerate the particles effectively. Such turbulence may occur with the cloud collapse phenomena. Since energy of the turbulence is comparable to cloud gravitational energy, it may significantly slow down the cloud collapse. Hence part of the gravitational energy may get transformed into turbulent energy of the GMCs. \cite{dogiel87} argued that if there existed a mechanism by which this turbulent energy was transformed into the particle energy, then there would be an effective acceleration inside the GMCs. Although this mechanism is theoretically possible as suggested by \cite{dogiel87}, it has not been observed yet. We assume that the conditions for reacceleration of CRs  are satisfied in a small number of GMCs.
 

In order to apply this mechanism in our work, we have selected the GMCs carefully. In general, we applied three necessary conditions to select 7 GMCs from \cite{chen20}, in which we assumed reacceleration of the particles is happening. We will show that a small contribution from these GMCs is enough to explain the positron excess adequately. The conditions for selection are the following,

1) The B parameter, defined above, must be less than 0.2. It has been mentioned in \cite{aharonian16} that in the cases of sources of angular extensions smaller than 1$^{\circ}$, \textit{Fermi}-LAT is able to detect the GMCs with B $\geq$ 0.4. In case where the compact GMCs are located in uncrowded region, \textit{Fermi}-LAT detection threshold can be as small as B $\approx$ 0.2. Again for very close GMCs (d $<<$ 1 kpc), the detection threshold can exceed B $\approx$ 1 to compensate the loss of the sensitivity of \textit{Fermi}-LAT due to large extensions of the GMCs. Just to ensure the fact that these GMCs are always outside the detection threshold of \textit{Fermi}-LAT as of now, we select only the GMCs for which B $<$ 0.2. 
Thus this condition takes into account of the fact that there may be GMCs in which reacceleration is happening but they are not detected by \textit{Fermi}-LAT. 

2) Next we select only the GMCs within 1 kpc of the Earth. Since leptons lose energy very efficiently by radiative losses, it is necessary to select nearby GMCs, so that they contribute significantly to the observed lepton spectrum. Reacceleration can happen in far away GMCs too, but their contribution will not be significant. 

3) As mentioned in \cite{dogiel87}, the turbulence occuring in the GMCs has a scale length. If the size of the GMCs is less than the maximum scale length of the plasma turbulence, then the particles will escape the GMCs, before getting reaccelerated properly. The maximum scale length of plasma turbulence inside a GMCs is of the order of $\sim$ 10 pc \cite{dogiel87}. That is why we selected only the GMCs which have radius greater than 10 pc. In this way, we take into account the fact that the particles are getting reaccelerated before escaping from the GMCs.

\begin{table*}[htbp]
\centering
\caption{\label{tab4} 7 selected GMCs parameters : Galactic coordinates (l, b), Radius, distances from the Earth (d), Mass and the B parameter from \cite{chen20}.}

\begin{tabular}{ccccccc}
\hline
GMC ID  & l & b & Radius & d & Mass & B \\
   & (deg) & (deg) & (pc) & (pc) & ($M_\odot$)  &\\
\hline
27 & 121.498 & -7.378 & 15.027 & 983.0$\pm$23.2 & 12313.7 & 0.12 \\
233 & -147.177 & -9.806 & 10.750 & 866.7$\pm$20.5 & 8483.8 & 0.11 \\
286 & 174.871 & 5.824 & 11.969 & 982.0$\pm$23.2 & 14307.2 & 0.14\\
288 & -75.048 & 7.070 & 12.095 & 911.5$\pm$21.5 & 8152.1  & 0.098\\
295 & 47.013 & -5.399 & 11.434 & 914.5$\pm$21.6 & 9601.4 & 0.11\\
342 & -96.318 & 4.705 & 11.041 & 952.7$\pm$22.5 & 14479.9 & 0.15\\
385 & 142.282 & 1.032 & 11.859 & 967.4$\pm$39.6 & 12580.0 & 0.13\\
\hline
\end{tabular}

\end{table*}   

Based on these three conditions, we have selected 7 GMCs, in which we assume reacceleration is happening. The physical description of these GMCs are listed in table \ref{tab4}. There is no other GMC which passes through these selection criteria in our set of GMCs.

{It has been shown in \cite{dogiel87}, that due to the effect of reacceleration, the spectrum of injected  particles gets harder. It has been calculated that the spectral index of the injected positron is -1.7, which is indicative of hard spectrum. We use this hardened spectra from 7 selected GMCs, and show that even if a very small flux of positrons is injected with this hardened spectrum, then taking their contribution into account, the positron excess can be explained. Next, we give the equations needed to calculate the flux of leptons after traversing the distance from the nearby GMCs to the Earth.}  

The magnetic field inside the GMCs is higher compared to the mean interstellar magnetic field \cite{crutcher12}. 
The secondary electrons and positrons produced  in nearby GMCs are expected to lose energy before they are injected into the ISM. The radiative loss of higher energy leptons is more than the lower energy ones, as a result we expect an exponential cut-off in their spectrum at high energy. Also strong gradients may be present in the CR distribution inside the GMCs, which may enhance the generation of plasma waves and thus, suppress the diffusion coefficient (\cite{gabici07}, \cite{gabici09}). This suppressed diffusion coefficient inside the GMCs will lead to suppression of secondary leptons that will actually get injected into ISM, because not all of the secondary leptons produced will be able to escape the GMC environment due to diffusive confinement and get injected in the ISM. So, in order to take this realistic situation into account, we have assumed that not all, but a majority fraction ($\sim$ 90 $\%$) of the total secondary lepton spectra produced, will be injected from the GMCs into the ISM. This suppression of secondary leptons will reflect in the normalisation of injected lepton spectra considered below.

 The injection spectra is expressed as a power law in Lorentz factor of the injected electrons and positrons $\gamma_e=E_e/m_e c^2$,
\begin{equation}
\label{eq12}
\begin{split}
Q(\gamma_e, d) &= Q_0\gamma_{e}^{-\beta_{e}}\:exp\left(-\frac{\gamma_e}{\gamma_{e,c}}\right)\delta(d)\\			
\end{split}
\end{equation}    
where the cut-off Lorentz factor $\gamma_{e,c}=E_{e,c}/m_e c^2$, unit of $Q_0$ is $\text{GeV}^{-1}\text{s}^{-1}$, d is the distance of each cloud from the observer and the Dirac delta function in this case signifies that we are considering point sources. During propagation in the ISM for time scales ($t$) less than $10^7$ years, the dominant radiative loss processes of relativistic electrons and positrons are synchrotron and inverse Compton (IC) scattering. 
The formalism for including the propagation effects by solving the transport equation including radiative losses and diffusion has been discussed in \cite{atoyan95}.
 The expression for IC and synchrotron energy loss term $p_2$ has been used from \cite{atoyan95},
\begin{equation}
\label{eq13}
\begin{split}
p_2 &= 5.2\times10^{-20}\frac{w_0}{1\:\frac{eV}{cm^{3}}} s^{-1}
\end{split}
\end{equation}
where $w_0 = w_B + w_{MBR} + w_{opt}$, $w_B$ is the energy density of the magnetic field, $w_{MBR}$ is microwave background radiation energy density, $w_{opt}$ is energy density of optical-IR radiation in interstellar space. For our study, we assume $w_0 \approx 1 \frac{eV}{cm^{3}}$.

The diffusion term has been included following \cite{atoyan95},
\begin{equation}
\label{eq14}
\begin{split}
D(\gamma_e) &= D_0\:\left(1+\left(\frac{\gamma_e}{\gamma_{e,*}}\right)\right)^\delta
\end{split}
\end{equation} 
Thus $D$ is constant for $\gamma_e << \gamma_{e,*}$, and energy dependent  for $\gamma_e \geq \gamma_{e,*}$, where $\gamma_{e,*}=E_{e,*}/m_e c^2$.

 
For point sources emitting continuously with a constant rate during the time $0\:\le\:t'\le\:t$, we get the following energy spectrum,

\begin{equation}
\label{eq15}
\begin{split}
f_{st}(d, t, \gamma_e) &= \frac{Q_0\gamma_e^{-\beta_e}}{4\pi\:D(\gamma_e)\:d} \ {\erfc}\left(\frac{d}{2\sqrt{D(\gamma_e)t_{\gamma_{e}}}}\right)\\
&\rm\:\:\:\:\:\:\:\:\:\:\:\:\:\:\:\:\:\:\:\:\:\:\:\:\:\:\:\:\:\:\:\:\:\:\times\:exp\left(-\frac{\gamma_e}{\gamma_{e,c}}\right)
\\
\end{split}
\end{equation}
where $t_{\gamma_{e}} = min(t, \frac{a}{p_2\gamma_e})$ and a = 0.75 \cite{atoyan95}, in our case $t>>\frac{a}{p_2\gamma_e}$.

The electron and positron flux from nearby GMCs {without} including solar modulation effect is,
\begin{equation}
\label{eq16}
\begin{split}
J^{e^\pm}_{obs} (\gamma_e)&= \left(\frac{c}{4\pi}\right)f_{st}(d,t,\gamma_e)
\end{split}
\end{equation}
{While calculating total observed spectrum from these nearby GMCs on Earth, we also take into account solar modulation effect, described by equation \eqref{eq7}}. The values of the relevant parameters used to calculate the secondary electron and positron fluxes from nearby {GMCs Taurus, Lupus, Orion A }and {7 selected GMCs} are listed in table \ref{tab6}. Using equation \eqref{eq16}, we calculate the total $e^{\pm}$ fluxes from these 10 nearby GMCs, which is our CASE 3. 

Finally, we add the lepton fluxes from CASE 1, CASE 2 and CASE 3 in order to fit the observational data. {It must be noted that secondary particles such as antiprotons, are also produced in these nearby GMCs. But we have checked using the formalism used in \cite{joshi15}, that the antiproton flux contribution from Taurus, Lupus, Orion A and the 7 selected GMCs, is very less. 
So we take into account these 10 GMCs in Histogram 1, as previously mentioned, while calculating different hadronic spectra/ratios as well as antiproton flux, so that however negligible their contribution might be, they get consistently included in the process of producing secondary CRs. Secondary Borons produced in these GMCs are taken into account in the same way. Leptons lose energy radiatively very fast, thus nearby GMCs contribute more significantly compared to the GMCs which are far away from the Earth. Thus these nearby GMCs are modelled individually. Taurus, Lupus and Orion A are the nearest GMCs analysed in detail in the work by \cite{aharonian16}. Also since we are considering 7 GMCs from \cite{chen20} as possible reacceleration sites, those GMCs are also modelled separately. In the next section, we discuss the results we have got from our simulated model.}

\section{Results}
\label{sec:4}
{In this section, we discuss the results that we have got from our model. We have divided our results in four subsections. In the first subsection we discuss the simulated proton, antiproton fluxes and B/C, $^{10}$Be/$^9$Be ratios. In the second subsection, we discuss the lepton fluxes and positron fraction and show how our model fits the positron spectrum. In the third subsection, we discuss lepton dipole anisotropy from the nearby GMCs. { In the last subsection, we discuss the uncertainties in CR propagation model parameters and expected fluxes from the nearby GMCs.} We have used the plain diffusion (PD) model to study CR propagation with the \textbf{\textbf{DRAGON}} code assuming the sources follow SNR distribution \cite{ferriere01}. This model includes diffusion and interactions of CRs {but the effect of reacceleration or convection were not considered}.}
\subsection{Protons / Cosmic ray nuclei / Antiprotons}
\label{subsec:4.1}

\begin{figure}[t]
\centering 
\includegraphics[width=.49\textwidth,origin=c,angle=0]{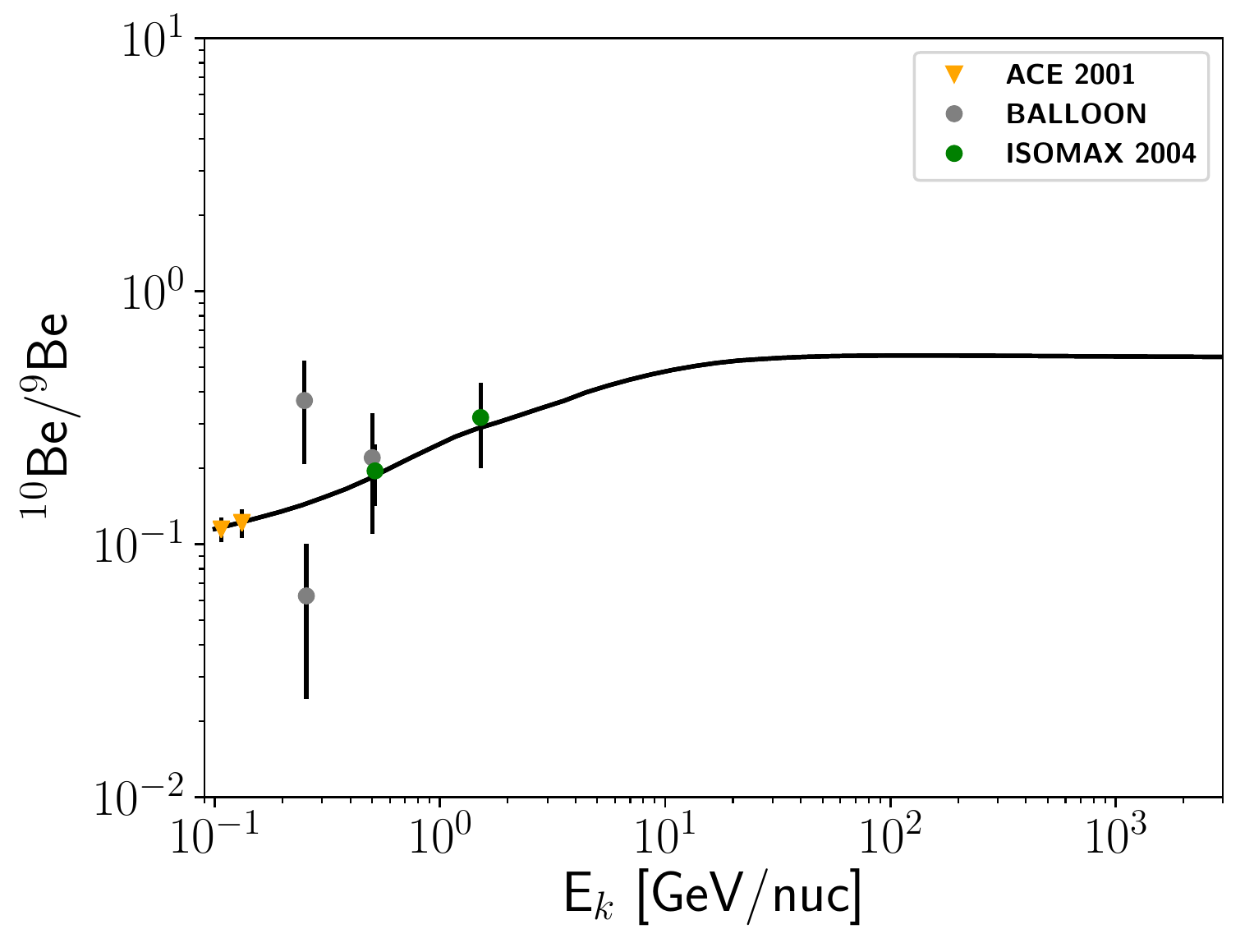}\caption{\label{fig5} $^{10}$Be/$^9$Be ratio calculated using \textbf{\textbf{DRAGON}} code, and plotted with the observational data given by 
ACE data \cite{ACE01}, several Balloon data and ISOMAX \cite{ISOMAX04} data. The black line signifies the simulated value of the ratio. The solar modulation potential was considered to be $\phi$ = 0.2 GV.}
\end{figure}
\begin{figure}[t]
\centering 
\includegraphics[width=.49\textwidth,origin=c,angle=0]{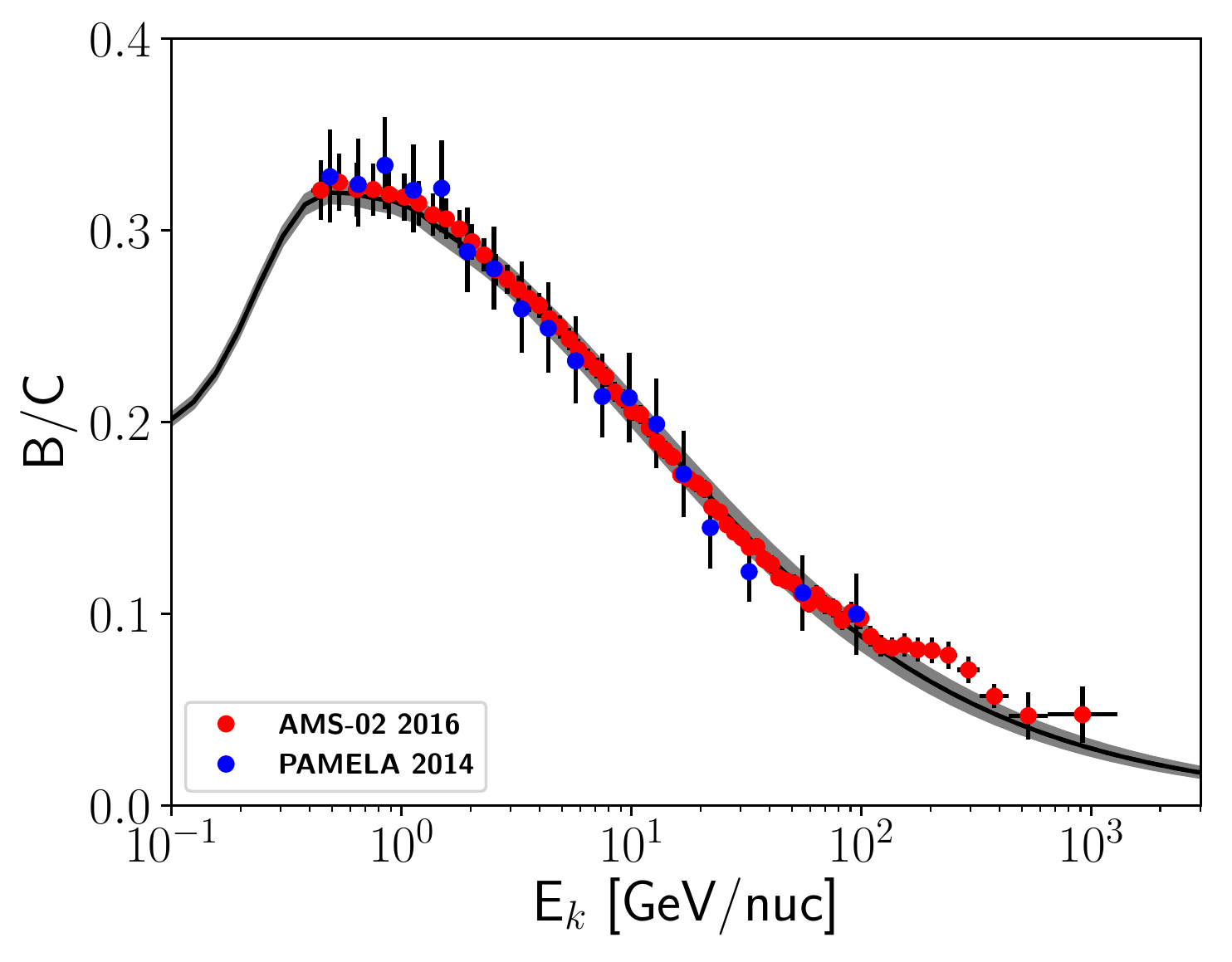}
\includegraphics[width=.49\textwidth,origin=c,angle=0]{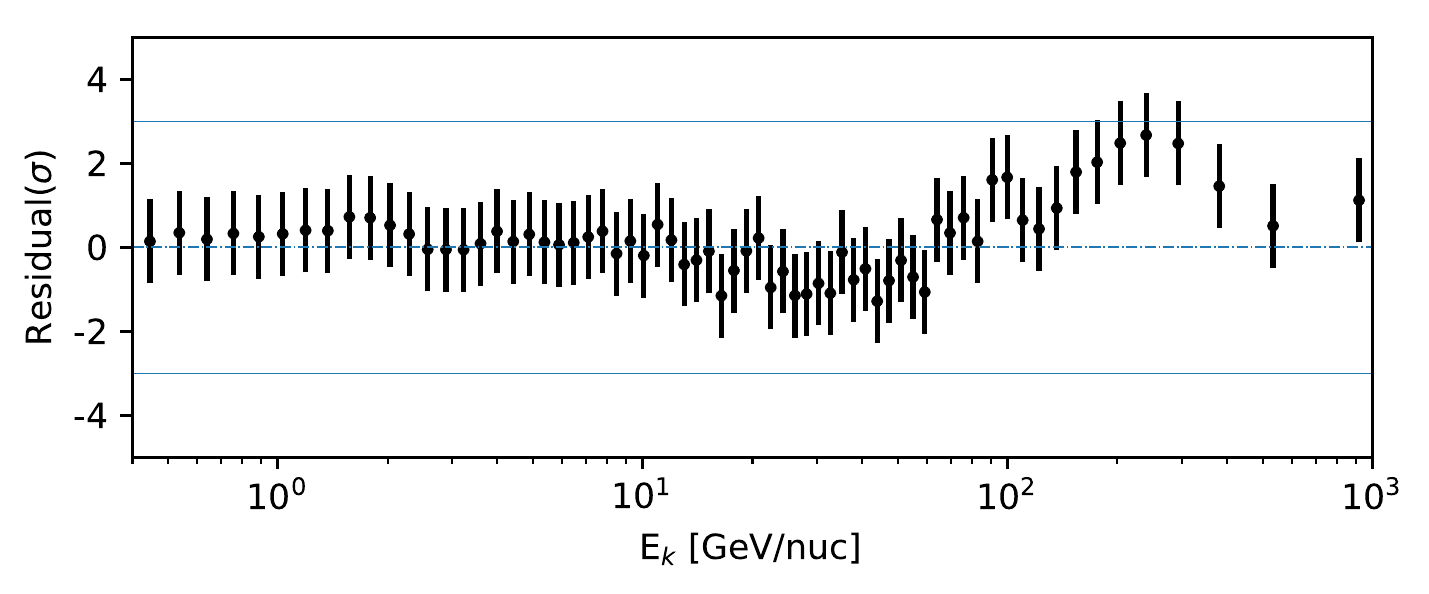}
\caption{\label{fig6} \textit {Upper panel:} B/C ratio plotted against the observational data reported by AMS-02 \cite{aguilar16a} and PAMELA \cite{adriani14}. Solid black line is the simulated ratio. The solar modulation potential was considered to be $\phi$ = 0.0 GV. {The gray shaded region signifies uncertainties due to variation of CR propagation parameters.} \textit{Lower panel:} Corresponding residual plot for the fit of simulated B/C ratio to the observed data. The solid blue line signifies 3$\sigma$ confidence level. The $\chi^2/\text{D.O.F.}$ for this fit of B/C ratio is $\approx$ 0.84.}
\end{figure}

\begin{table}[]
\caption{\label{tab5} Models and parameter values selected in PD model to fit the various observed CR nuclei spectra and ratio, using \textbf{\textbf{DRAGON}}, are listed in this table. The parameters used here have been discussed before. $D_0$ in this case is the normalisation of the diffusion coefficient used for (CASE 1 + CASE 2 + CASE 3). $v_A$ is the Alfven velocity, $v_w$ is the wind or convection velocity, $dv_w/dz$ is vertical gradient of convection velocity.}

\begin{tabular}{ccc}
\hline
Model/Parameter & Option/Value\\
\hline
$R_{max}$  &  25.0 kpc\\
$z_t$ & 8.0 kpc\\
L & 24.0 kpc\\
HI gas density type & \cite{gordon76,dickey90,cox86}\\
HII gas density type & \cite{cordes91}\\
$H_2$ gas density type & Equation \eqref{eq9}\\
Source distribution & Ferriere \cite{ferriere01} \\
Diffusion type & Exponential (see equation \eqref{eq2})\\
$D_0$ & 2.4 $\times 10^{29}\:cm^2/s$\\
$\rho_0$ & 4.0 GV\\
$\delta$ & 0.53 \\
$\eta$ & -\:0.40\\
$v_A$ & 0\\
$v_w$ & 0\\
$\frac{dv_w}{dz}$ & 0\\
Magnetic field type & Pshirkov \cite{pshirkov11}\\
$B_0^{disk}$ & 2$\times 10^{-6}$ Gauss\\
$B_0^{halo}$ & 4$\times 10^{-6}$ Gauss\\
$B_0^{turbulent}$ & 6.1$\times 10^{-6}$ Gauss (see equation \eqref{eq1})\\
$\alpha_1^k/\alpha_2^k$ & 1.95/2.33\\
$\rho_{br, 1}^k$ & 7\:GV\\
\hline

\end{tabular}

\end{table}

\begin{figure}[t]
\centering 
\includegraphics[width=.49\textwidth,origin=c,angle=0]{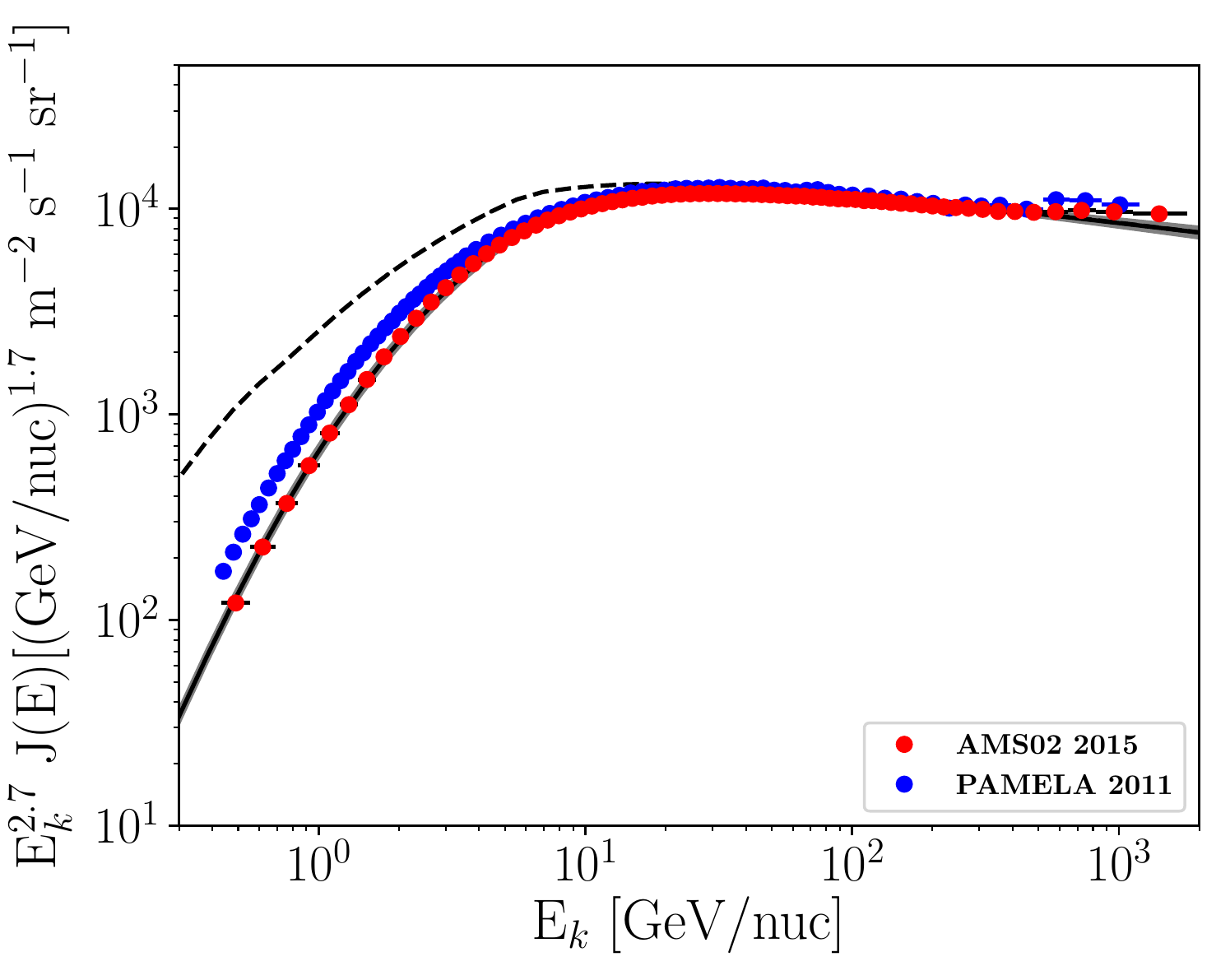}
\includegraphics[width=.49\textwidth,origin=c,angle=0]{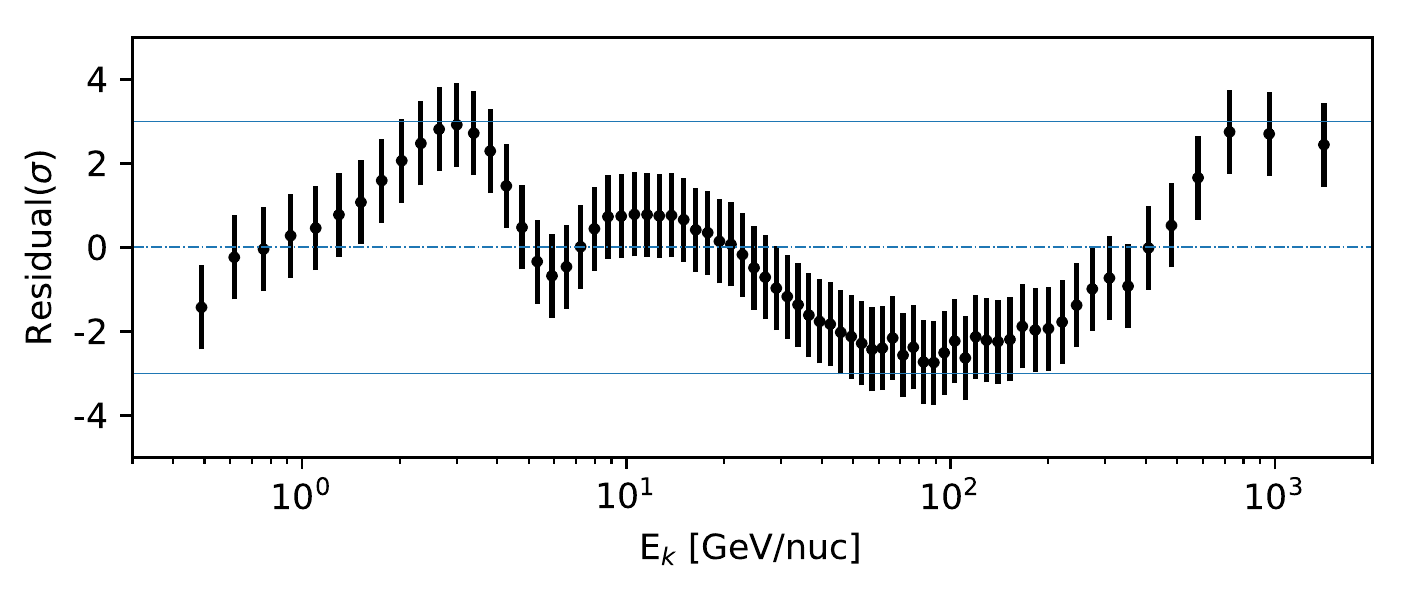}
\caption{\label{fig7} \textit{Upper panel:} Proton flux calculated using \textbf{\textbf{DRAGON}} code, and plotted with the observational data given by AMS-02 \cite{aguilar15a} and PAMELA \cite{adriani11dataa}. The solar modulation potential considered is $\phi$ = 0.564 GV. The solid (dashed) black line corresponds to solar modulated (unmodulated) proton spectrum. {The gray shaded region signifies uncertainties due to variation of CR propagation parameters.} \textit{Lower panel:} Corresponding residual plot for the fit of simulated proton spectrum to the observed data. The solid blue line signifies 3$\sigma$ confidence level. The $\chi^2/\text{D.O.F.}$ for this fit of proton spectrum $\approx$ 2.9.}
\end{figure}
\begin{figure}[htbp]
\centering 
\includegraphics[width=.49\textwidth,origin=c,angle=0]{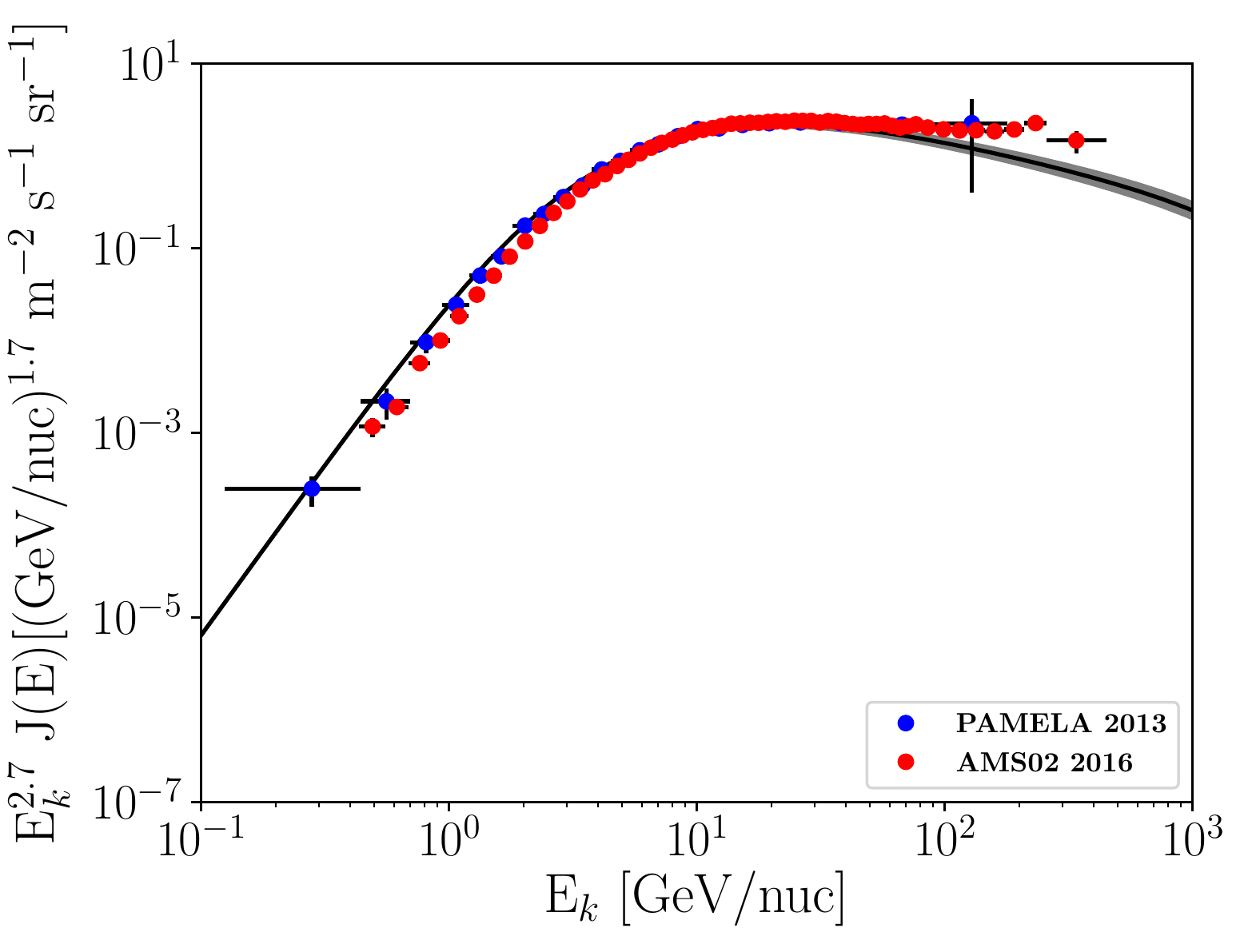}
\caption{\label{fig8} \textit{Upper panel:} Antiproton flux calculated using \textbf{\textbf{DRAGON}} code, and plotted with the observational data given by AMS-02 \cite{aguilar16b} and PAMELA \cite{adriani13datab}. The solar modulation potential considered is $\phi$ = 0.564 GV. The solid black line corresponds to solar modulated antiproton spectrum. {The gray shaded region signifies uncertainties due to variation of CR propagation parameters.}}
\end{figure}

\begin{table*}[htbp]
\centering
\caption{\label{tab6} Table for the parameters used to calculate total $e^{\pm}$ flux observed on Earth from {Taurus, Lupus, Orion A and 7 selected GMCs}. Parameters: $Q_0$ is injection normalisation, $\beta_e$ is the spectral index for $e^{\pm}$ injection from the GMCs, $D_0$ is the diffusion coefficient normalisation, $\delta$ is the diffusion index, $E_{e,*}$ is reference energy for diffusion coefficient, $E_{e,c}$ is the cutoff energy, $\phi$ is the solar modulation potential. $D_0$ in this case, is the diffusion coefficient normalisation used for CASE 3.}

\begin{tabular}{cccccccc}
\hline
Cloud  & $Q_0$ & $\beta_e$ & $D_0$ &  $\delta$ & $E_{e,*}$ & $E_{e,c}$ & $\phi$ \\
 & ($\text{GeV}^{-1}s^{-1}$) & & ($\text{cm}^2/\text{s}$) &  & (GeV) & (GeV) & (GeV) \\
\hline
Taurus & 8.5$\times 10^{43}$ & 2.83 & 2.4$\times 10^{29}$ & 0.53 & 5$\times 10^3$ & 5$\times 10^3$ & 0.564 \\
Lupus & 1$\times 10^{43}$ & 2.72 & 2.4$\times 10^{29}$ & 0.53 &  5$\times 10^3$ & 5$\times 10^3$ & 0.564 \\
Orion A & 3.6$\times 10^{44}$ & 2.81 & 2.4$\times 10^{29}$ & 0.53 &  5$\times 10^3$ & 5$\times 10^3$ & 0.564\\

GMC ID 27 & 2.8$\times 10^{38}$ & 1.7 & 2.4$\times 10^{29}$ & 0.53 & 5$\times 10^3$ & 6.5$\times 10^2$ & 0.564\\
GMC ID 233 & 2.8$\times 10^{38}$ & 1.7 & 2.4$\times 10^{29}$ & 0.53 & 5$\times 10^3$ & 6.5$\times 10^2$ & 0.564\\
GMC ID 286 & 2.8$\times 10^{38}$ & 1.7 & 2.4$\times 10^{29}$ & 0.53 & 5$\times 10^3$ & 6.5$\times 10^2$ & 0.564\\
GMC ID 288 & 2.8$\times 10^{38}$ & 1.7 & 2.4$\times 10^{29}$ & 0.53 & 5$\times 10^3$ & 6.5$\times 10^2$ & 0.564\\
GMC ID 295 & 2.8$\times 10^{38}$ & 1.7 & 2.4$\times 10^{29}$ & 0.53 & 5$\times 10^3$ & 6.5$\times 10^2$ & 0.564\\
GMC ID 342 & 2.8$\times 10^{38}$ & 1.7 & 2.4$\times 10^{29}$ & 0.53 & 5$\times 10^3$ & 6.5$\times 10^2$ & 0.564\\
GMC ID 385 & 2.8$\times 10^{38}$ & 1.7 & 2.4$\times 10^{29}$ & 0.53 & 5$\times 10^3$ & 6.5$\times 10^2$ & 0.564\\
\hline
\end{tabular}

\end{table*} 

 In our analysis, we have fitted the observed CR nuclei data in the following way. {We have used the standard model given by \cite{gordon76,dickey90,cox86} for neutral and atomic hydrogen, and the model given by \cite{cordes91} for the ionized hydrogen. Using these models, we have modelled the ISM hydrogen gas distribution (CASE 1). For molecular hydrogen gas distribution, we have included all the GMCs considered in our work (CASE 2 + CASE 3). We have used the fit parameters of Histogram 1 discussed previously, to take into account molecular hydrogen content of all of the GMCs. Then we have implemented equation \eqref{eq9} to model the molecular hydrogen gas density distribution in our simulation.}

We have fixed the halo height $z_t$ to a {value for which the $^{10}$Be/$^9$Be data is well fitted. Such ratios of unstable to stable isotope of secondary particles is a major constraint for CR propagation. $^{10}$Be is the unstable isotope of berylliam, which is unstable to $\beta$ decay. Since being unstable, $^{10}$Be decays faster than its stable counterpart after getting produced from CR interactions inside the Galaxy. The decay time of $^{10}$Be becomes longer than the escape time from the Galactic halo for rigidity above 10-100 GV, depending on the size of the halo z$_t$. This is why measurement of $^{10}$Be/$^9$Be ratio is sensitive to the paramater z$_t$. Recently \cite{evoli20} has shown that optimum halo height must be around $\approx$ 7 kpc. That is why we take the halo height considered in our work around that value. We take 8 kpc as our halo height and fit the observed $^{10}$Be/$^9$Be ratio. The fit for $^{10}$Be/$^9$Be ratio is shown in figure \ref{fig5}, and from there it can be seen that the choice of our halo height is a good fit for the observed ratio.}

We have then subsequently set $B_0^{turbulent}$ using equation \eqref{eq1}. 
{Next we have estimated the average diffusion coefficient in the Galaxy by fitting the B/C observed data. Boron is secondary produced from spallation of Carbon, which is primary in nature. If the halo height is larger, then the secondary borons are  produced more and also they spend longer time in the Galaxy, making the secondary to primary ratio larger. On the other hand, if the diffusion in the Galaxy is high, then Borons escape the Galaxy faster, and the ratio becomes smaller. Hence it can be seen that the B/C ratio scales with halo height (z$_t$) and diffusion coefficient D as z$_t$/D.} We have fixed the value of the reference rigidity ($\rho_0$) and adjusted the normalisation ($D_0$), $\delta$ and $\eta$ to get a good fit to the observed data of B/C ratio. {The ratio and its corresponding residual are plotted in figure \ref{fig6}. From the figure and the residual plot, it can be seen that our estimation of the average diffusion coefficient is accurate.} 

Then the spectral indices and breaks of the injected CR spectra are adjusted to get a good fit to the observed proton data given by AMS-02 and PAMELA. {We use a low energy break at around 7 GV. No other high energy break is used to fit proton spectrum. The plot for proton spectrum and its corresponding residual plot are shown in figure \ref{fig7}. It can be seen that the residual calculated for each data point is confined within 3$\sigma$ confidence level.} 

{We also calculate secondary antiprotons produced in our model. We use CASE 1 + CASE 2 + CASE 3 together to estimate the total flux of antiprotons as a secondary product. Primary CRs interact with ISM gas and molecular hydrogen clumped inside GMCs, and produce antiprotons. By taking into account all of the cases together, we get the estimate of antiprotons produced in our model setup. The antiproton spectrum is shown in figure \ref{fig8} against the data obtained by AMS-02 and PAMELA. The same parameters in table \ref{tab5} were used to find the total antiproton spectrum.}

\subsection{Leptons}
\label{subsec:4.2}

We have adjusted the injection spectrum of primary electrons in our model, given by broken power-law, to get a good fit to the observed data by AMS-02 and PAMELA. {The conventional primary electron sources follow the Ferriere distribution \cite{ferriere01}.} The CR protons and heavy nuclei injected, also interact with neutral and ionised hydrogen gas in ISM (CASE 1) and molecular hydrogen in GMCs in the Galactic plane (CASE 2), producing secondary electrons and positrons.
 
\begin{figure}
\centering 
\includegraphics[width=.49\textwidth,origin=c,angle=0]{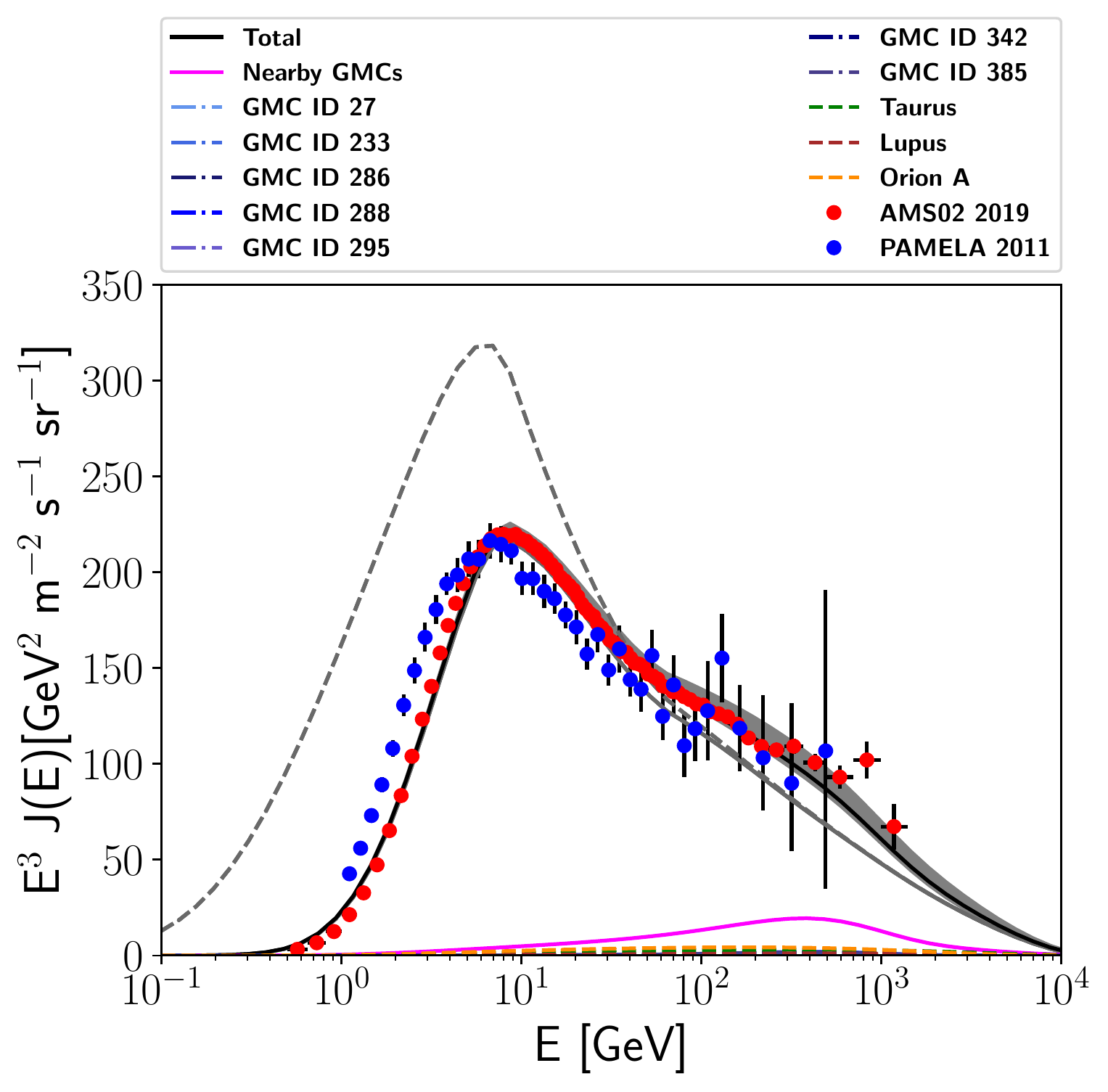}
\includegraphics[width=.49\textwidth,origin=c,angle=0]{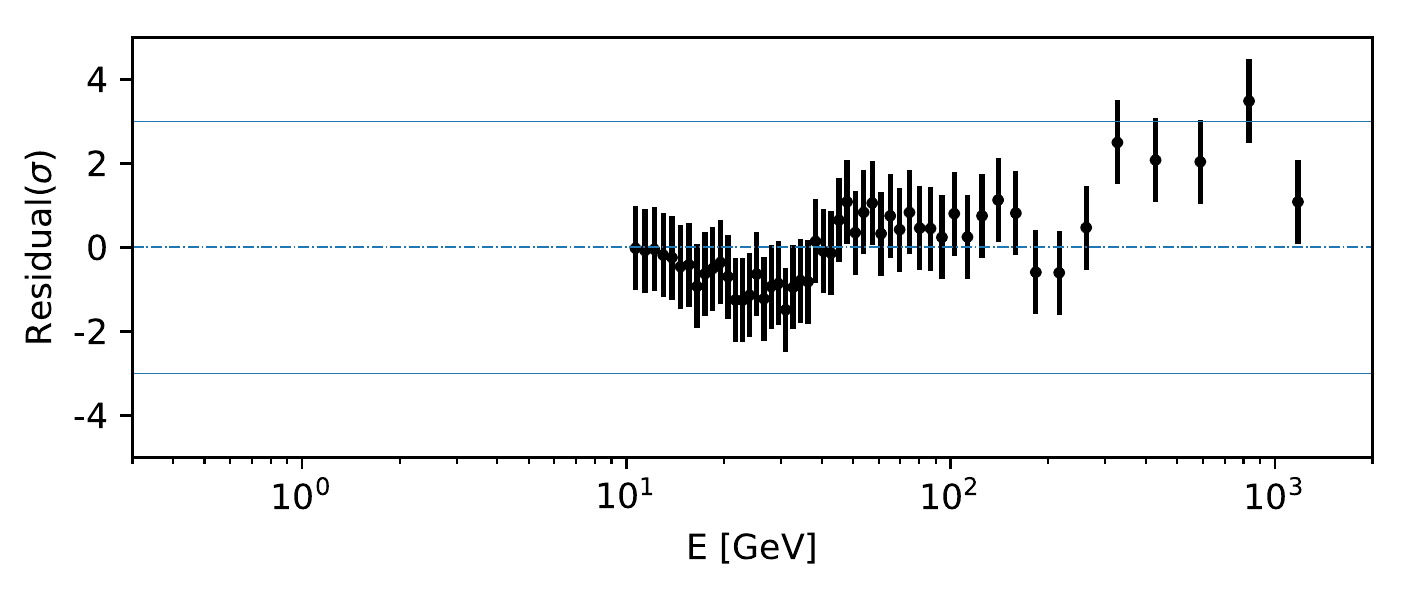}
\caption{\label{fig9} \textit{Upper panel:} Electron flux calculated using \textbf{\textbf{DRAGON}} code, and plotted with the observational data given by AMS-02 \cite{aguilar19a} and PAMELA \cite{adriani11datab}. Solid (dashed) gray line is the solar modulated (unmodulated) total flux for (CASE 1 + CASE 2). Magenta line shows the total flux from nearby GMCs (CASE 3).  Black line corresponds to the total flux calculated from our work. The solar modulation potential considered is $\phi$ = 0.564 GV. {The gray shaded region signifies uncertainties due to variation of CR propagation parameters and uncertainties in the normalisation of nearby GMC fluxes.}
\textit{Lower panel:} Corresponding residual plot for the fit of simulated electron spectrum to the observed data from 10 GeV energy and above. The solid blue line signifies 3$\sigma$ confidence level. {The $\chi^2/\text{D.O.F.}$ for this fit of electron spectrum above 10 GeV is $\approx$ 1.22.}}
\end{figure}

\begin{figure}
\centering 
\includegraphics[width=.49\textwidth,origin=c,angle=0]{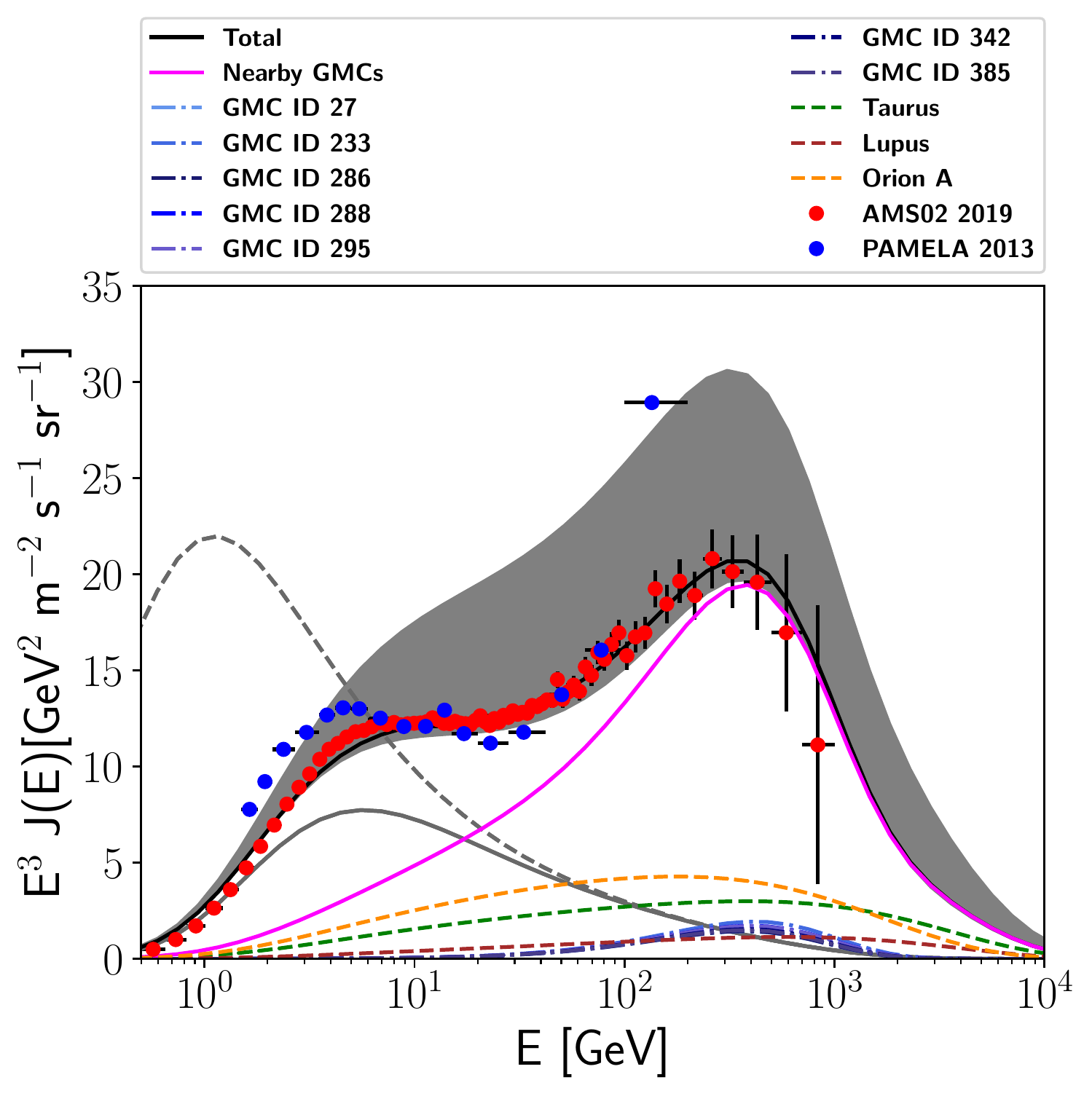}
\includegraphics[width=.49\textwidth,origin=c,angle=0]{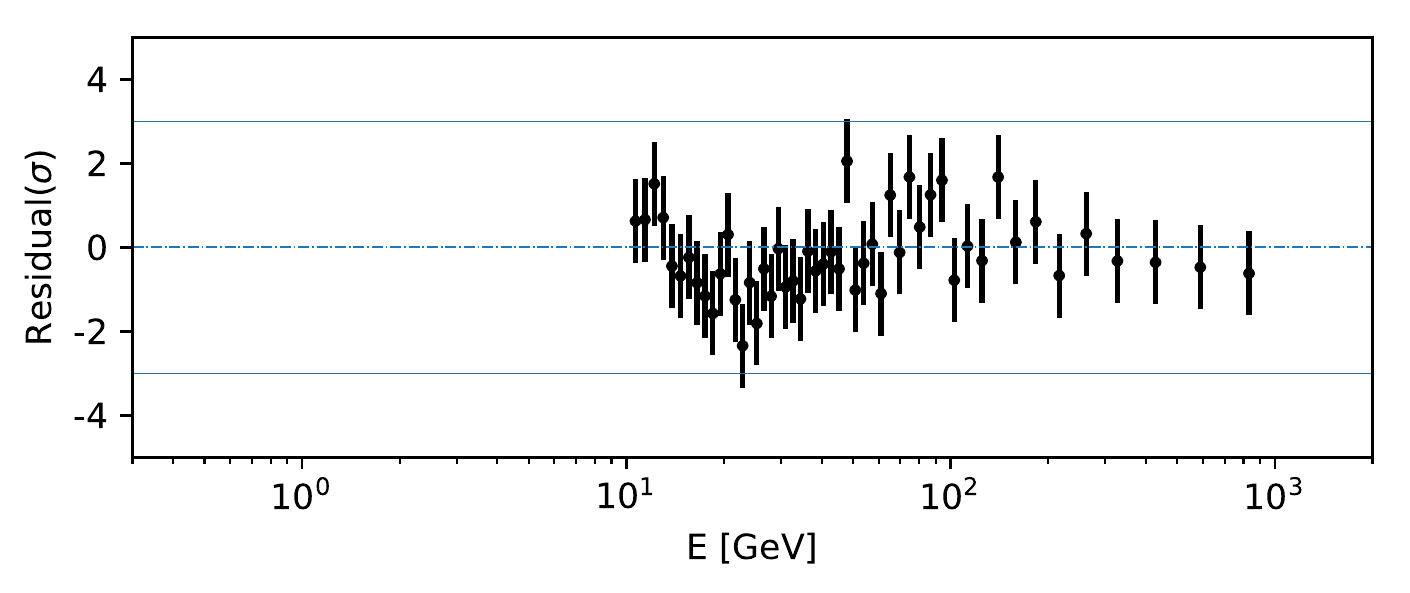}
\caption{\label{fig10} \textit{Upper panel:} Positron flux using \textbf{\textbf{DRAGON}} code, and plotted against the observational data reported by AMS-02 \cite{aguilar19b} and PAMELA \cite{adriani13dataa}. Solid (dashed) gray line is the solar modulated (unmodulated) total flux for (CASE 1 + CASE 2). Magenta line shows the total flux from nearby GMCs (CASE 3).  Black line corresponds to the total flux calculated from our work.  The solar modulation potential considered is $\phi$ = 0.564 GV. {The gray shaded region signifies uncertainties due to variation of CR propagation parameters and uncertainties in the normalisation of nearby GMC fluxes.}
\textit{Lower panel:} Corresponding residual plot for the fit of simulated positron spectrum to the observed data from 10 GeV energy and above. The solid blue line signifies 3$\sigma$ confidence level. {The $\chi^2/\text{D.O.F.}$ for this fit of positron spectrum above 10 GeV is $\approx$ 0.96.}}
\end{figure}

\begin{figure}[t]
\centering 
\includegraphics[width=.49\textwidth,origin=c,angle=0]{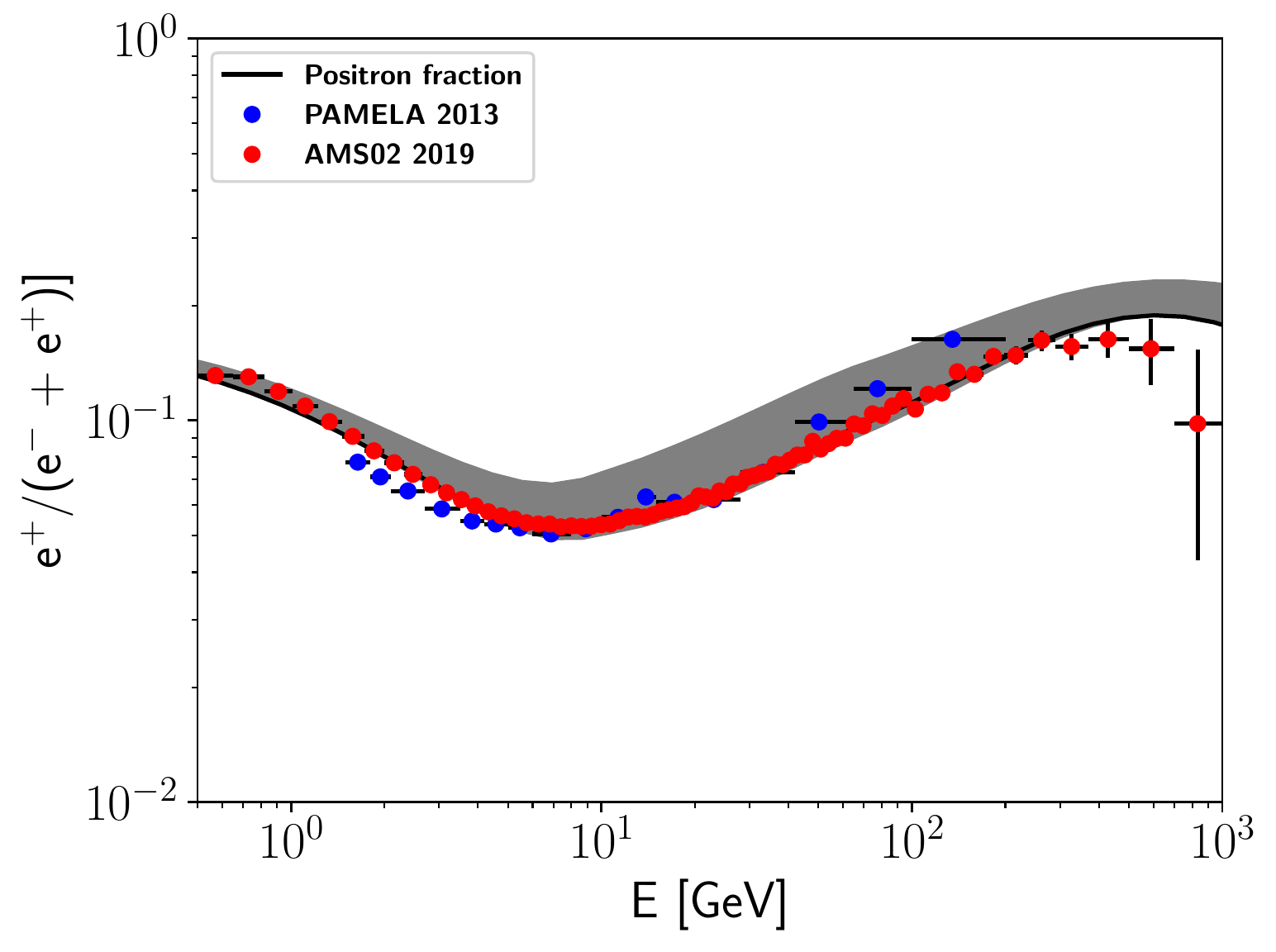}
\includegraphics[width=.49\textwidth,origin=c,angle=0]{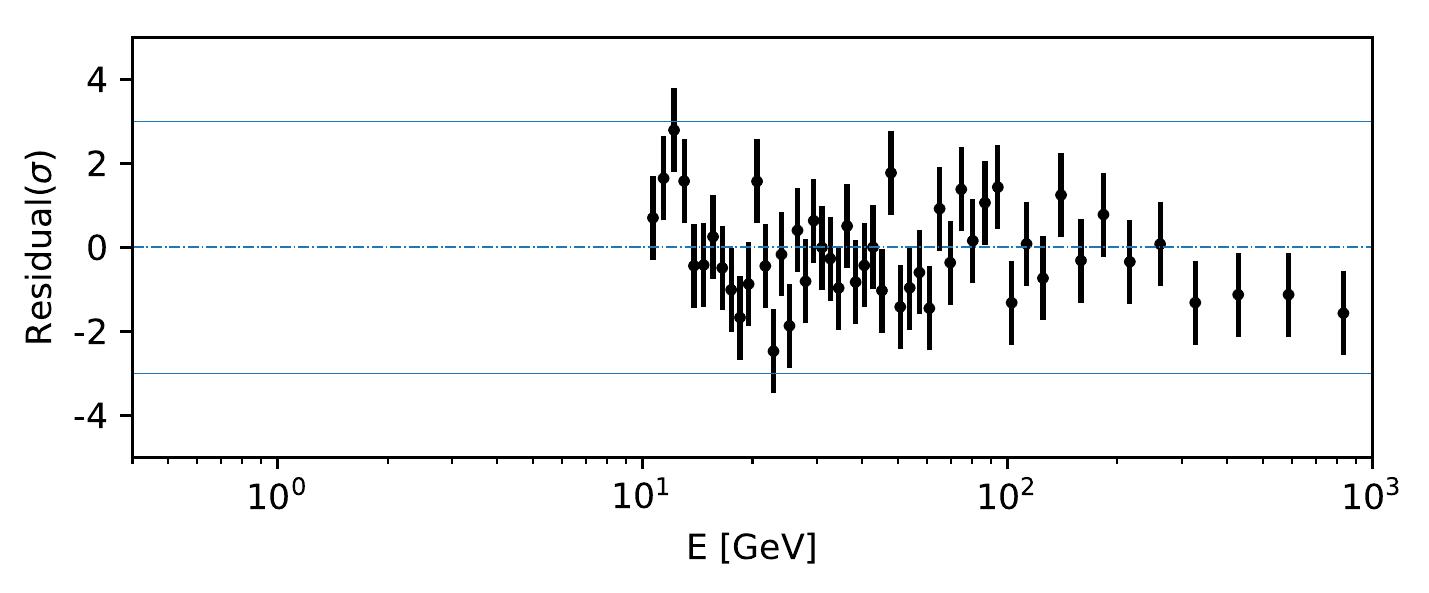}
\
\caption{\label{fig11} \textit{Upper panel:} Positron fraction calculated using \textbf{\textbf{DRAGON}} code, and plotted with the observational data given by AMS-02 \cite{aguilar19b} and PAMELA \cite{adriani13dataa}. {The gray shaded region signifies uncertainties due to variation of CR propagation parameters and uncertainties in the normalisaton of nearby GMC fluxes.}
\textit{Lower panel:} Corresponding residual plot for the fit of simulated positron fraction to the observed data from 10 GeV energy and above. The solid blue line signifies 3$\sigma$ confidence level.}
\end{figure} 
 
 We have used the catalog of GMCs near the Galactic plane from \cite{rice16}, {\cite{chen20}, and  \cite{aharonian16}. We omit out 7 GMCs from \cite{chen20} based on the selection criteria mentioned in the previous section. Also we omit Taurus, Lupus and Orion A from \cite{aharonian16}, and take the contribution of other GMCs into account in CASE 2}. We have taken bin size of 100 pc in radial distance to build the histograms for the number of GMCs in each bin. {We have taken all the GMCs of CASE 2 to build the number histogram, which can be fitted by a linear combination of Gaussian distribution function and Lorentzian distribution function}. These distributions are used with average number density of hydrogen molecules $<n_{H_2}>\:\sim 10\:\:cm^{-3}$ {to get the molecular hydrogen density distribution in the Galaxy considered in CASE 2}. Instead of discrete clumps of GMCs, we have assumed continuous distribution of matter along the Galactic plane, due to this reason the number density of hydrogen molecules is lower than typical density of hydrogen molecules in GMCs  \cite{heyer15}.
 
\begin{table}[t]
\caption{\label{tab7}Models and parameter values selected in PD model to fit the observed lepton spectra and positron fraction, using \textbf{\textbf{DRAGON}}, are listed in this table. The parameters used here have been discussed before. $D_0$ in this case is the diffusion coefficient normalisation used for (CASE 1 + CASE 2). $v_A$ is the Alfven velocity, $v_w$ is the wind or convection velocity, $dv_w/dz$ is vertical gradient of convection velocity.}

\begin{tabular}{ccc}
\hline
Model/Parameter & Option/Value\\
\hline
$R_{max}$  &  25.0 kpc\\
$z_t$ & 8.0 kpc\\
L & 24.0 kpc\\
HI gas density type & \cite{gordon76,dickey90,cox86}\\
HII gas density type & \cite{cordes91}\\
$H_2$ gas density type & Equation \eqref{eq9}\\
Source distribution & Ferriere \cite{ferriere01} \\
Diffusion type & Exponential (see equation \eqref{eq2})\\
$D_0$ & 2.4 $\times 10^{29}\:cm^2/s$\\
$\rho_0$ & 4.0 GV\\
$\delta$ & 0.53 \\
$\eta$ & -\:0.40\\
$v_A$ & 0\\
$v_w$ & 0\\
$\frac{dv_w}{dz}$ & 0\\
Magnetic field type & Pshirkov \cite{pshirkov11}\\
$B_0^{disk}$ & 2$\times 10^{-6}$ Gauss\\
$B_0^{halo}$ & 4$\times 10^{-6}$ Gauss\\
$B_0^{turbulent}$ & 6.1$\times 10^{-6}$ Gauss (see equation \eqref{eq1})\\
$\alpha_1^{e}/\alpha_2^{e}/\alpha_3^{e}$ & 2.0/2.7/2.4\\
$\rho_{br, 1}^{e}/\rho_{br, 2}^{e}$ & 8/65\:\: GV\\
$\rho_c^{e}$ & 10 TeV\\
\hline
\end{tabular}

\end{table}
 
Electron spectrum is dominated by primary CR electrons, {which are produced in SNRs that follow Ferriere distribution. Contribution from primary electron sources, along with the contribution of secondary electrons produced in interactions of CRs with ISM gas,} are included in our CASE 1.
{The contribution of secondary electrons produced from primary CRs interacting with GMCs considered in Histogram 2, are included in our CASE 2}. {Positrons are also produced as secondaries in interactions between primary CR nuclei and atomic, ionized component of hydrogen in the ISM (CASE 1) and also, molecular hydrogen component distributed in the Galaxy (CASE 2).} We add up the contributions from CASE 1 and CASE 2 to fit the electron and positron spectra. {The combined contribution of CASE 1 + CASE 2 is shown with gray line in the plots of electron and positron spectra. The parameters used for simulating CASE 1 + CASE 2 for electron and positron spectra are presented in tables \ref{tab5} and \ref{tab7}.} 

It is obvious that CR positrons require more nearby sources to explain the observed data. Hence, we consider the contribution of CR interactions in GMCs close to the Earth, which {have been analyzed with \textit{Fermi}-LAT data in the work by \cite{aharonian16}}. We have considered three GMCs, Taurus, Lupus and  Orion A, which are closest to the Earth, {for which gamma ray data have been analyzed}. {Other GMCs which have been studied with  \textit{Fermi}-LAT data in the work by \cite{aharonian16}, are too far from the Earth to be able to contribute significantly. Due to this reason we have included them in Histogram 2 of CASE 2, and did not model them individually in CASE 3.} 
{Further, we assume there is reacceleration of CRs due to magnetized turbulence in a few nearby GMCs. Due to this process, the CR spectrum gets hardened inside these GMCs.  The secondary leptons  produced in CR interactions inside these GMCs also have a hard spectrum. The spectral index of this hardened spectrum comes out to be -1.7, which has been calculated in the work by \cite{dogiel87}. We consider such hardened injection spectrum from 7 GMCs selected from \cite{chen20}, based on three criteria, which have been discussed in the previous section. The total contribution from Taurus, Lupus, Orion A and a small contribution from these 7 GMCs, due to reacceleration, is considered as CASE 3 in our model. The values of the necessary parameters of these GMCs required to fit the data are given in table \ref{tab6}.}

{In figure \ref{fig9} and \ref{fig10}, the electron flux and positron flux are shown against the data given by AMS-02 and PAMELA. In figure \ref{fig11}, corresponding positron fraction is also shown. The residuals for these plots are also shown. Since below 10 GeV, heliospheric modulation plays an important role, and positron spectrum effectively starts to rise from 10 GeV, we show the residuals from 10 GeV and above. Data points below 10 GeV were not used while plotting the residuals for electron, positron spectra, and positron fraction. {Please note the differences between the spectral shapes of Taurus, Lupus, Orion A and 7 selected GMCs. These differences arise due to the differences between the parameters for the source term of nearby GMCs. For Taurus, Lupus and Orion A, the injected secondary lepton spectra is soft in nature \cite{aharonian16}, whereas for 7 selected GMCs, we get a harder injected secondary lepton spectra \cite{dogiel87}, because of the assumption of reacceleration happening inside these GMCs due to magnetized turbulence. Hence, there are differences in the spectral indices of injected lepton spectra $\beta_e$. The cutoff energy $E_{e,c}$ is also different for these 7 GMCs. In 7 GMCs where reacceleration due to magnetized turbulence is assumed, it can be expected that the magnetic field is higher than that of average GMCs, otherwise every other GMCs would have shown signature of reacceleration in their respective gamma ray analysis. Since the average magnetic field is higher in these 7 GMCs, compared to that of Taurus, Lupus and Orion A, radiative losses are more due to the synchrotron process and thus, the cut-off in spectrum is expected at comparatively lower energy. This makes the cutoff energy ($E_{e,c}$) of the 7 selected GMCs lower compared to that of Taurus, Lupus and Orion A. See table \ref{tab6} for the distinction between the injected spectral indices $\beta_e$ and cutoff energy $E_{e,c}$ of these GMCs.}}

\subsection{Anisotropy due to nearby GMCs}
\label{subsec:4.3}
{Nearby electron-positron sources can induce anisotropy that can be observed on Earth. This anisotropy is mainly determined by structure of the magnetic field in the solar neighbourhood, which can be calculated by the formalism given in \cite{berezinskii90}, as $\frac{3D}{v}$ $\frac{\Delta N}{N}$. Here v is the relativistic speed of the CRs, and D is the diffusion coefficient for effective collision frequency $V^2/D$ of CRs. For nearby GMCs, the anisotropy can be calculated by \cite{berezinskii90,grassoaniso09,joshi17}}

\begin{equation}
\label{eq17}
\begin{split}
Anisotropy\:(\delta) = \frac{3\:d}{2\:c\:t_{\gamma_{e}}}\:\frac{N^{GMC}_{e^- + e^+}}{N^{total}_{e^- + e^+}}
\\
\end{split}
\end{equation}\\

\begin{figure}[t]
\centering 
\includegraphics[width=.49\textwidth,origin=c,angle=0]{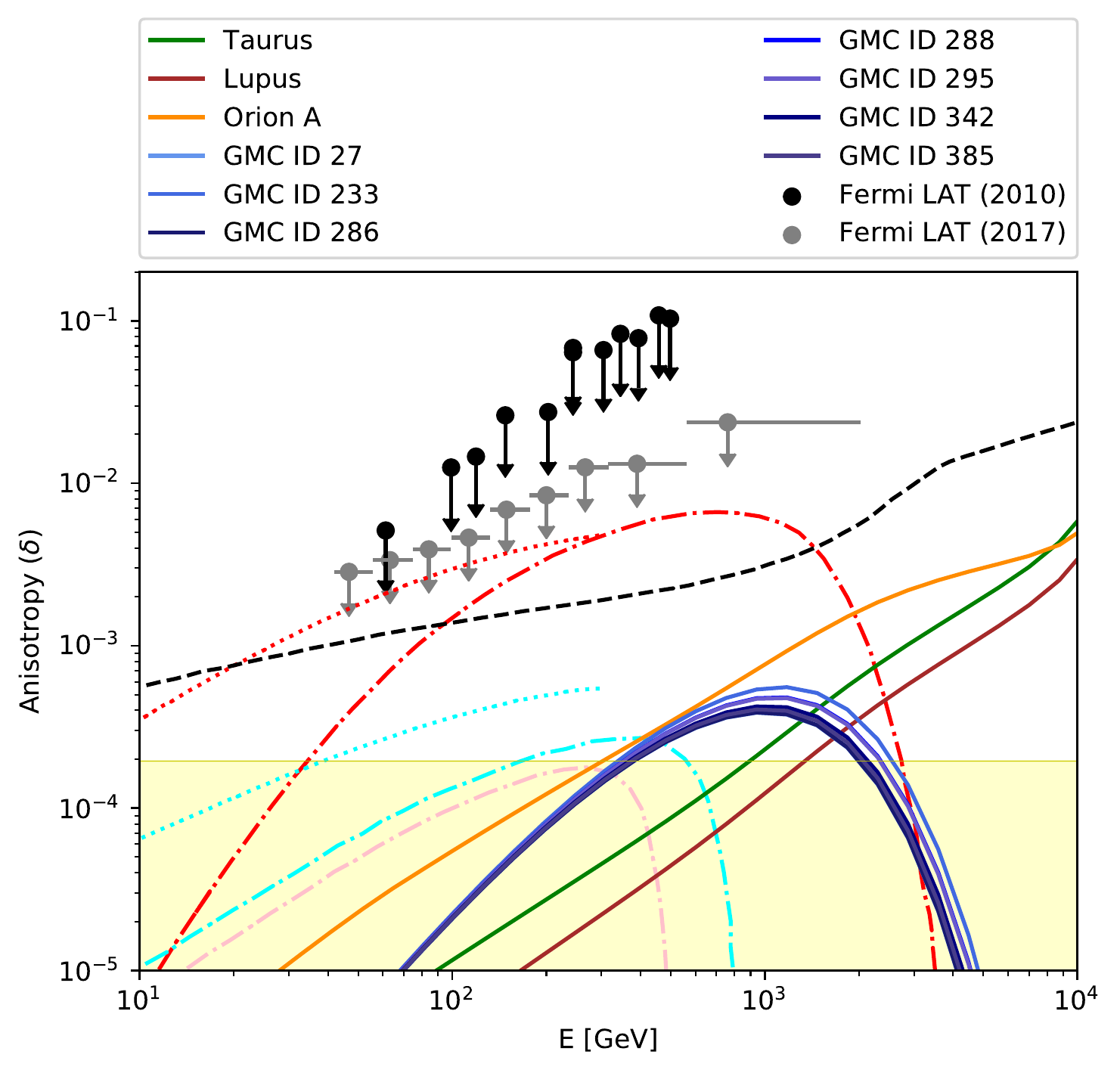}
\
\caption{\label{fig12} Cosmic ray electron + positron anisotropy for nearby GMCs (CASE 3) in comparison with \textit{Fermi}-LAT upper limits. The upper limits are taken from \cite{ackermann10} and \cite{abdollahi17}. { We also show electron + positron  anisotropy generated from other candidates for positron excess, pulsars and dark matter. The black dashed line is the anisotropy calculated for the dark matter distributed in Milky Way Galaxy \cite{ackermann10}. The dot-dashed lines are for astrophysical objects such as pulsars, Monogem (red), Geminga (cyan) and B1055-52 (pink) \cite{joshi17}. The dotted lines are also for pulsars, but taken from the work of \cite{hooper09}(see figure 5 from \cite{hooper09}). The red dotted line is for Monogem (B0656+14) like source and cyan dotted line is for Geminga like source. The yellow filled region signifies the $\delta$ threshold at the 2$\sigma$ level, for $\dot{N_{ev}}$ = 3 $\times$ $10^7$ electrons and positrons per year above 10 GeV, and $t_{obs}$ = 7 years. Above this region, Fermi gamma ray telescope should be able to detect dipole anisotropies at 2$\sigma$ confidence level \cite{hooper09}.}}
\end{figure}

{Here d is the distance of nearby GMCs from the Earth, $t_{\gamma_{e}}$ is the energy loss timescale due to IC and synchrotron processes. The $e^{\pm}$ pair emission ratio $\text{N}^{GMC}_{e^- + e^+}/\text{N}^{total}_{e^- + e^+}$ from all sources observed on Earth, determines the nearby discrete source anisotropy. In figure \ref{fig12}, we show the anisotropy calculated for the nearby GMCs considered in our work, against the upper limits by \textit{Fermi}-LAT (\cite{ackermann10}, \cite{abdollahi17}). Note that the anisotropy in lepton spectra from Taurus, Lupus and Orion A increases with energy because of their soft secondary lepton spectra, while due to the hard secondary lepton spectra from the 7 GMCs of \cite{chen20}, the anisotropy decreases sharply with energy.}

\subsection{Uncertainties in Propagation Model Parameters and Fluxes from GMCs}
\label{subsec:4.4}
{In this subsection, we discuss the uncertainties that were considered in our work in model parameters of CR propagation and expected fluxes from nearby GMCs (Taurus, Lupus and Orion A). 

First we have performed a systematic $\chi^2$ analysis around the best-fit values for the fit of B/C ratio. The standard $\chi^2$ analysis is given by,

\begin{equation}
\label{eq18}
\begin{split}
\chi^2 = \sum\limits_{i=1}^n \left[\frac{y_i^{obs}(E) - y_i^{mod}(E, a_M)}{\sigma_i}\right]^2
\\
\end{split}
\end{equation}\\

where, $y_i^{obs}(E)$ is the B/C ratio data observed by AMS-02, $y_i^{mod}(E, a_M)$ is the simulated values of our model at specific energies respectively, and $a_M$ are the values of M parameters in the simulation. The standard error of each observed value is given by $\sigma_i$.  

As mentioned earlier, the fit for B/C ratio depends on halo height ($z_t$) and parameters of diffusion ($D_0$, $\delta$). We have varied the CR propagation parameters $z_t$, $D_0$ and $\delta$ around their best-fit values. In order to find the allowed range of these parameters, we have restricted the value of the reduced $\chi^2$ to be less than 3 for the B/C data, similar to the treatment that has been done in \cite{lio05}. The minimum and maximum values for these three parameters, for which the reduced $\chi^2$ values are within 3, are given in table \ref{tab8}. The effect of the uncertainties in these CR propagation parameters are shown with a gray region in the B/C ratio plot in figure \ref{fig6}. Also, the effect of this variation of CR propagation parameters on proton spectrum and antiproton spectrum are shown with gray regions in figures \ref{fig7} and \ref{fig8} respectively.

We consider the uncertainties in these propagation parameters while calculating the lepton spectra and the positron fraction. We also consider the uncertainties in the normalisation of the parent proton flux inside the nearby GMCs Taurus, Lupus and Orion A. The uncertainties in the parent proton spectra inside these three GMCs are given in table \ref{tab3} of \cite{aharonian16} from \textit{Fermi}-LAT observations of gamma-ray spectra. The uncertainties in these parent proton spectra lead to uncertainties in the injected secondary lepton spectra from these three GMCs. We have considered a fraction ($\sim$ 90 $\%$) of the upper and lower limits of uncertainties in the normalisation of the injected lepton spectra from Taurus, Lupus and Orion A, along with the uncertainties in the CR propagation parameters to estimate the total uncertainties in the electron, positron spectrum and the positron fraction. The total uncertainty in the electron spectrum, positron spectrum and positron fraction are shown with gray regions in figure \ref{fig9}, \ref{fig10} and \ref{fig11} respectively. The maximum and minimum values of the normalisation of the injected lepton spectra (considering the suppression) from Taurus, Lupus and Orion A, due to uncertainty in the parent proton spectra, are given in table \ref{tab8}.

The only free parameter in our model is the normalisation of the injected lepton spectra from the 7 selected GMCs, taken from the catalog of \cite{chen20}. As mentioned before, we have assumed reacceleration due to magnetized turbulence inside these GMCs. Since the phenomena of reacceleration  occuring inside the GMCs is yet to be observed by \textit{Fermi}-LAT, we can not constrain this normalisation parameter from the observed gamma-ray data. But we can get an idea about this parameter if we assume that the luminosity of each of the 7 selected GMCs is comparable to the luminosity of Taurus or Lupus or Orion A. 
The lepton luminosity of the GMCs is related to the normalisation constant $Q_0$ in $\text{GeV}^{-1}s^{-1}$ through the relation,
\begin{equation}
\label{eq19}
\begin{split}
L_{e,GMC} = \int_{1\:\text{GeV}}^{10^4\:\text{GeV}} E_e\:Q(\gamma_e, d)\:dE_e
\\
\end{split}
\end{equation}\\
 in unit of $\text{GeV}\:\text{s}^{-1}$. 
 The order of magnitude of the luminosity in leptons of each of the 7 GMCs in our work is the same as that of Taurus or Lupus, and it is one order of magnitude lower than that of Orion A. Hence the value  of normalisation $Q_0$ used in our work (see table \ref{tab6}) for the 7 selected GMCs is not unphysical and it gives the best fit to the observed positron data. This is why while calculating uncertainties, this normalisation was fixed at the best fit value. Future observations, and possible detection of reacceleration inside GMCs may give stronger constraints on the injection parameters of such GMCs.}

\begin{table*}
\caption{\label{tab8} Allowed values for the CR propagation parameters and normalisation of nearby GMCs for our model.}
\centering

\begin{tabular}{ccccccc}
\hline
Par./Val. & $z_t$ & $D_0$ & $\delta$ & $Q_0^{Taurus}$ & $Q_0^{Lupus}$ & $Q_0^{Orion A}$\\
& (kpc) & ($\text{cm}^2/\text{s}$) &  & ($\text{GeV}^{-1}s^{-1}$) & ($\text{GeV}^{-1}s^{-1}$) & ($\text{GeV}^{-1}s^{-1}$) \\
\hline
Minimum Value & 7 & 2.2 $\times$ 10$^{29}$ & 0.51 & 8.5$\times 10^{43}$ & 1$\times 10^{43}$ & 3.6$\times 10^{44}$\\
Best-fit Value & 8 & 2.4 $\times$ 10$^{29}$ & 0.53 & 8.5$\times 10^{43}$ & 1$\times 10^{43}$ & 3.6$\times 10^{44}$\\
Maximum Value & 9 & 2.6 $\times$ 10$^{29}$ & 0.55 & 1.7$\times 10^{44}$ & 2.1$\times 10^{43}$ & 7$\times 10^{44}$\\
\hline
\end{tabular}

\end{table*}

\section{Discussion}
\label{sec:5}

\subsection{Summary}
\label{subsec:5.1}

{In this work, we have comprehensively discussed the origin of the most prominent features in the electron and positron spectra, as seen by AMS-02 and PAMELA data in a different light compared to the existing literature. Supernova remnants in the Galaxy were considered to be the sites for acceleration of primary CRs. Primary CRs accelerated in SNRs are injected into the ISM and then these primary CRs interact with ISM hydrogen gas to produce secondary CRs. 
GMCs scattered in the Galactic plane, were considered to be major sites for secondary particle production from CR interactions. Primary CRs interact with cold protons inside  GMCs, and produce secondary CRs such as leptons, antiprotons and gamma-rays. In our paper, we show that total contribution from nearby GMCs (CASE 3), along with contributions from ISM (CASE 1) and all the other reported GMCs (CASE 2) can explain the positron flux very well.

First, we build a CR transport scenario using publicly available code {\bf \textbf{DRAGON}}, while considering CASE 1, CASE 2 and CASE 3. Using that CR transport setup, we reproduce $^{10}$Be/$^{9}$Be ratio, B/C ratio and proton spectrum. We show our simulated spectra and ratio against the data points provided by  AMS-02 and PAMELA. We also show the corresponding residuals in our model with respect to the observed data for each fit. Residual is defined by the ``distance" between simulated value and observed data, divided by total experimental error. As it can be seen from the residual plots, residuals are always confined within 3$\sigma$, confirming a good accuracy of the fitting \cite{evoli20}. We also compare the secondary antiproton spectrum produced from the interactions of primary CRs with the intervening medium (CASE 1 + CASE 2 + CASE 3) in our model, with the recent data by AMS-02 and PAMELA.

Next we consider the electron and positron data observed by AMS-02 and PAMELA. Positron data shows a rise above 10 GeV, and then a fall at around 200 to 300 GeV. We simulate CASE 1 and CASE 2 with {\bf \textbf{DRAGON}}, and model CASE 3 analytically. We consider that primary electrons are injected from SNRs with a broken power law injection spectrum. Also secondary electrons and positrons are produced from interactions of primary CRs with ISM gas (CASE 1) and GMCs considered in CASE 2. Next we calculate the contribution from nearby GMCs separately using the formalism given by \cite{atoyan95}. We take 3 nearest GMCs, for which \textit{Fermi}-LAT analysis were performed by \cite{aharonian16}. Leptons are produced inside these GMCs through pp interactions. Apart from this, we select 7 GMCs from the work by \cite{chen20}, in which reacceleration due to magnetized turbulence was considered. We refer to the work done by \cite{dogiel87}, and consider a small hardened component from these GMCs, which is due to reacceleration. Finally combining all of the above contributions, we show that the electron spectra, positron spectra and positron fraction can be well fitted by our model. The rise of the observed positron spectrum and subsequent fall can be well explained by the positron spectrum simulated in our model. Also, the corresponding residuals are shown in the figures. Since we are concerned with the positron excess phenomena, which is dominant from 10 GeV and above, and moreover, below 10 GeV, heliospheric modulation can alter the spectrum, we show the residuals of lepton fluxes and positron fraction, from 10 GeV and above, and neglect the residuals calculated below 10 GeV.
Finally we show the electron and positron anisotropy induced by nearby GMCs considered in our work. We have calculated the electron and positron anisotropy following the formalism given by \cite{grassoaniso09, joshi17}, and plotted the calculated anisotropy against the upper limits obtained by \textit{Fermi}-LAT. It can be seen that the anisotropy induced by the nearby GMCs considered in our work is lower than the upper limits provided by \textit{Fermi}-LAT, hence our model is very much plausible in terms of anisotropy signal. The positions of these nearby GMCs considered in our model, have been shown in figure \ref{fig13} and \ref{fig14}.}

\subsection{Distinguishing between different models in terms of anisotropy}
\label{subsec:5.2}

\begin{figure*}
\centering 
\includegraphics[width=1.\textwidth,origin=l,angle=0]{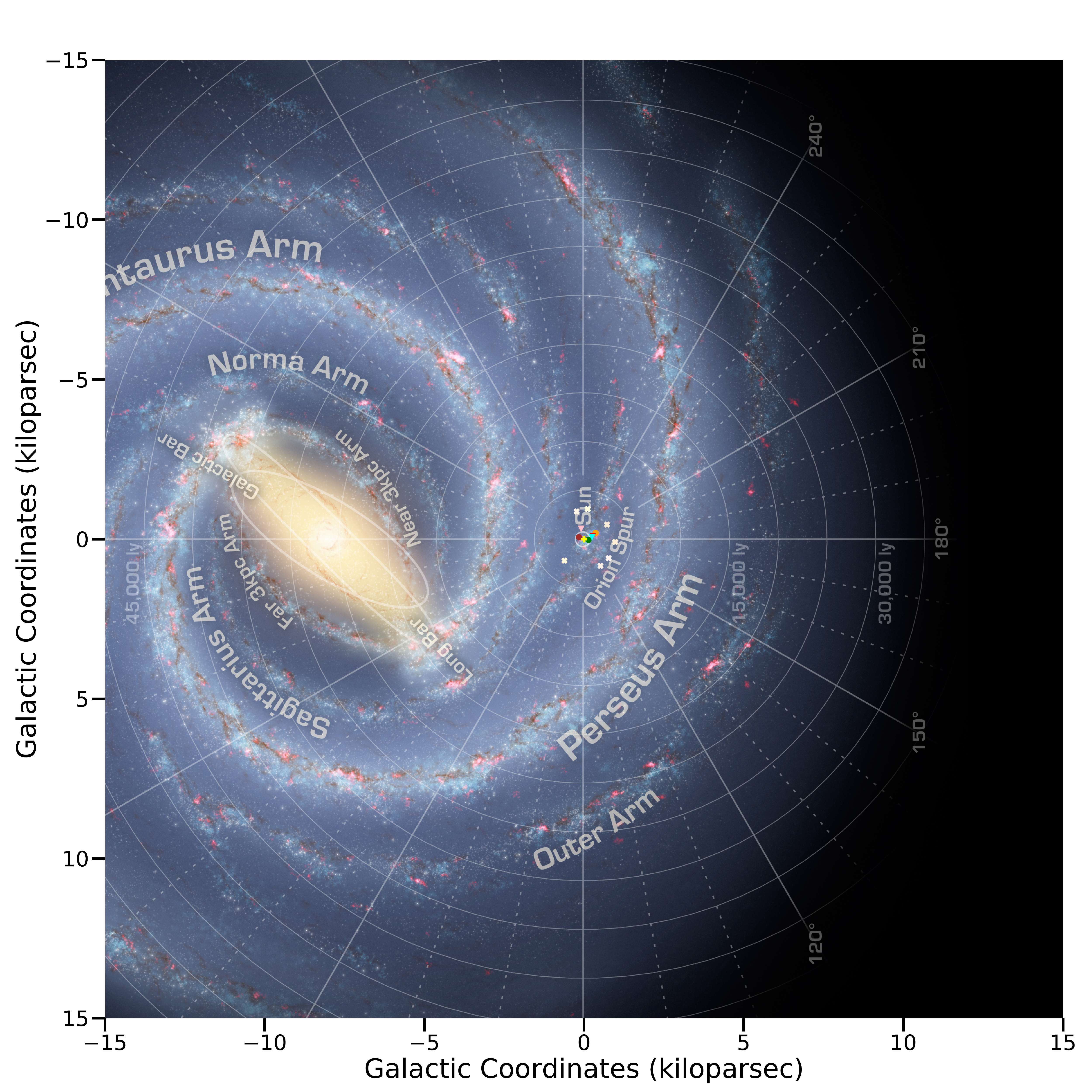}
\caption{\label{fig13} {GMCs considered in this work are plotted on the background of Milky Way Galaxy, alongwith the candidate pulsars, which are conventionally considered while explaining the positron excess. Background illustration \cite{churchwell09} produced by Robert Hurt of the Spitzer Science Center, reflecting the current understanding of Galactic structure. The color scheme is same as figure \ref{fig12}, other than 7 selected GMCs, where the color scheme is GMC ID 27 (linen), GMC ID 233 (antiquewhite), GMC ID 286 (papayawhip), GMC ID 288 (oldlace), GMC ID 295 (cornsilk), GMC ID 342 (lightyellow) and GMC ID 385 (seashell). The filled circles signify nearby GMCs, where pp interaction is considered (Taurus, Lupus and Orion A), cross marks signify 7 selected nearby GMCs where reacceleration is considered and filled triangles signify the nearby pulsars (Monogem, Geminga and B1055-52). The yellow plus mark is the position of the Sun.}}
\end{figure*}

\begin{figure}
\centering 
\includegraphics[width=.5\textwidth,origin=l,angle=0]{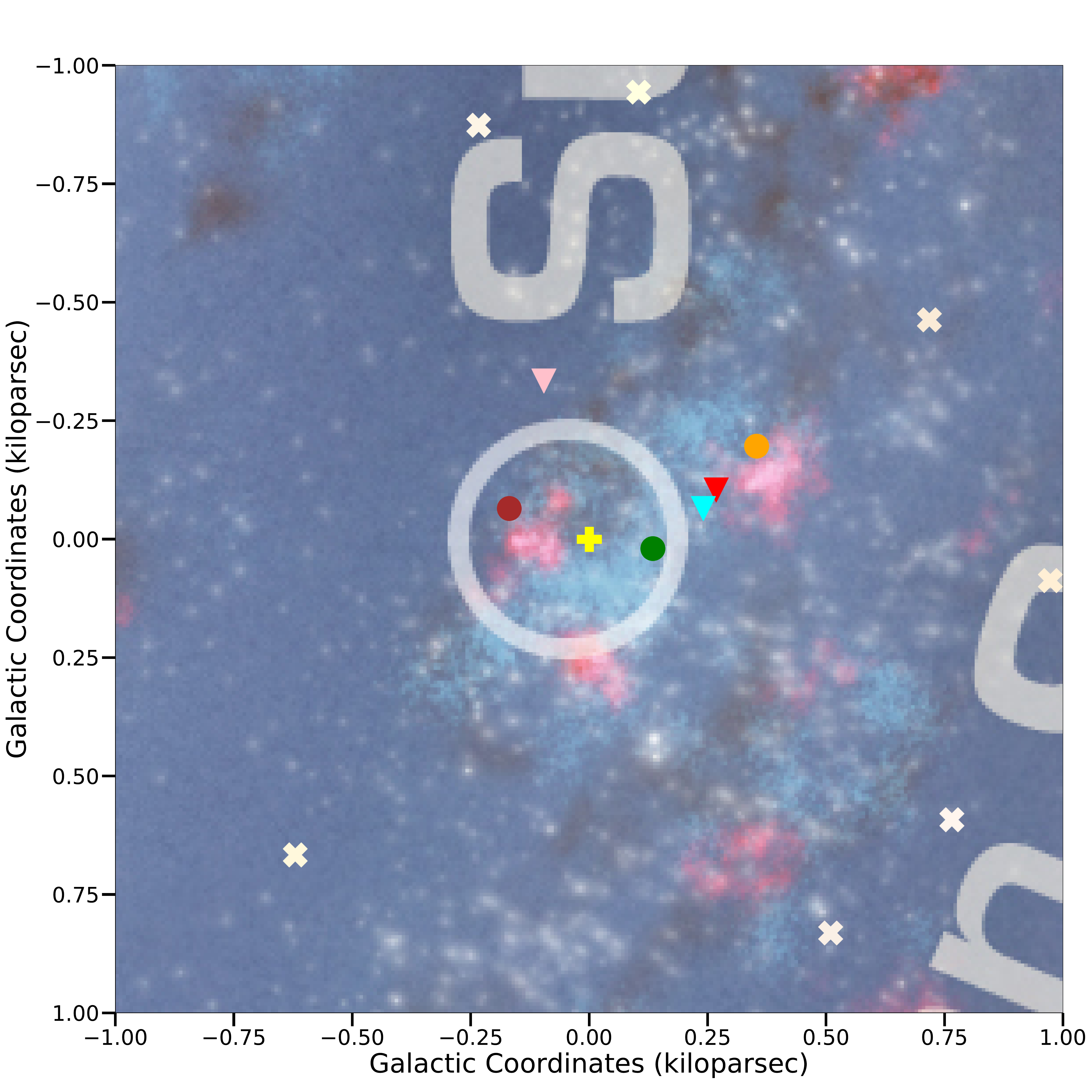}
\caption{\label{fig14} {Zoomed view of the region (radius of 1 kpc) around the Sun. The marker shapes and colors are same as described in figure \ref{fig13}. Background illustration \cite{churchwell09} produced by Robert Hurt of the Spitzer Science Center, reflecting the current understanding of Galactic structure.}}
\end{figure}

{As pointed out by \cite{hooper09}, fitting of the positron spectrum and positron fraction presented by AMS-02 and PAMELA alone, may be insufficient to distinguish between different scenarios considered for explaining this interesting phenomenon. As we have discussed earlier, dark matter distributed in the Galaxy and astrophysical objects such as pulsars are two conventional candidates for explaining the observed positron excess. However, in this work, we have presented an alternative model, which explains the positron excess using the secondary positrons produced from nearby and faraway GMCs distributed in the Milky Way Galaxy. Since explaining the positron excess through fitting of the positron spectrum and positron fraction using contribution from nearby GMCs is inadequate, we will discuss anisotropy signals as an additional measurement for solving this problem.

As discussed earlier, contrary to hadronic CRs, high energy CR electrons and positrons propagating through GMF lose energy rapidly through synchrotron radiation and IC collision with low energy photons of interstellar radiation field (ISRF). Thus, in order to contribute significantly to the positron spectrum, the contributing sources must be nearby, which will in turn induce anisotropy signals. Therefore, depending on the propagation properties in the GMF, detection of excess CR leptons, with energy sufficiently high enough to minimize geomagnetic field and heliospheric modulation effects will uncover the nature and the presence of such nearby CR sources \cite{ackermann10}. In other words, even after taking into account diffusion in the ISM, a dipole anisotropy should be present in the direction of dominant nearby sources at sufficiently high energies. Although anisotropy signal, which is not associated with nearby sources, can also be expected to result from Compton-Getting effect \cite{cg35}, where relative motion of the observer with respect to CR plasma changes the intensity of the CR fluxes, with larger intensity arriving from the direction of motion and lower intensity arriving from the opposite direction. But in general, anisotropy signal should be an useful probe to distinguish between several distinct models that are being used to explain the positron excess from the nearby sources. In particular, as \cite{hooper09} pointed out, that anisotropy at 2$\sigma$ level can be detected, if one fulfills the condition $\delta \gtrsim 2\sqrt{2}\:(\dot{N_{ev}}.t_{obs})^{-1/2}$, where $\dot{N_{ev}}$ is the rate of events detected per unit time above a given threshold and $t_{obs}$ is the observation time. In accordance with \cite{hooper09}, we have taken a rate of approximately 3$\times 10^7$ electrons per year above 10 GeV, and observation time of 7 years \cite{abdollahi17}. This implies that dipole anisotropy should get detected at the 2$\sigma$ confidence level in the electron-positron flux above 10 GeV if $\delta \gtrsim$ 0.02 $\%$. We show this threshold with a yellow region in figure \ref{fig12}. If anisotropies calculated from any candidates cross this threshold at sufficiently high energies, then it can be predicted that dipole anisotropies from those direction can be detected at 2$\sigma$ level. Otherwise anisotropies from the sources will not be significant enough to be detected and will be mixed with background isotropy due to diffusion. It can be readily seen from figure \ref{fig12}, that anisotropies calculated from all the nearby GMCs considered in this work, are well above this threshold. Hence our model predicts that in the GMC scenario proposed in this work, dipole anisotropy may be observed in the directions of the nearby GMCs considered.  

As shown by previous works, pulsars can be possible candidates for explaining positron excess. These nearby pulsars, specifically Monogem and Geminga can explain the positron spectrum, also their anisotropy signals are below anisotropy upper limit given \textit{Fermi}-LAT (\cite{ackermann10} and \cite{abdollahi17}). \cite{joshi17} has also shown that the pulsar B1055-52 can also be a nearby source that can contribute in the explanation of positron spectrum. It is shown in \cite{hooper09}, that a small dipole anisotropy may be observed by \textit{Fermi}-LAT at sufficiently high energies, in the direction of Monogem and Geminga. \cite{joshi17} has calculated anisotropy of Monogem, Geminga and B1055-52 pulsars in their respective directions, for an injection time of 85 kyr. Also since Monogem and Geminga lie in similar directions in the sky, they are expected to contribute the same overall anisotropy. The anisotropy calculated from the works by \cite{hooper09} and \cite{joshi17} have been shown in figure \ref{fig12}. The positions of these pulsars are shown in figures \ref{fig13} and \ref{fig14}. From figure \ref{fig12}, it can also be seen that Monogem and Geminga do induce a small dipole anisotropy in their respective direction at high energies. But anisotropy signal from pulsar B1055-52 can not be detected at 2$\sigma$ level, as it is below the anisotropy threshold. 

Alternatively, also annihilations or decay by Galactic dark matter distributed all over the Milky Way halo can also be primarily responsible for the positron excess observed. If this is to be the case, a dipole anisotropy must be generated towards the direction of Galactic center, since the dark matter is denser in that direction. A model anisotropy calculated from the dark matter distributed in the Galaxy \cite{ackermann10}, is shown in figure \ref{fig12}, which also is above the anisotropy threshold. Since, as seen from figure \ref{fig13}, apart from Taurus, every other GMCs considered in this work, and all the nearby  candidate pulsars are in different directions compared to the direction of Galactic Center, distinction can be made in terms of anisotropy between dark matter and GMCs and/or pulsars. Also by considering the difference in proximity and anisotropy signal between Taurus and dark matter residing in the Galactic center, it is possible to distinguish both of them. However, if there are any nearby subhalo of dark matter in the direction of posited GMCs and/or pulsars, then it would be hard to make any distinction among them in terms of anisotropy. Fortunately, the chance of nearby and luminous dark matter clumps that can explain the positron excess is very small for ordinary pair-annihilation cross section \cite{brun09}. For larger annihilation cross sections, the predicted associated gamma-ray flux from dark matter annihilation will exceed the point source sensitivity of \textit{Fermi}-LAT, i.e. it would have very likely been observed shining in gamma rays. In particular, \cite{profumo15} has shown that if an anisotropy from such a clump were detected, and if such anisotropy did not generate from anisotropic diffusion effects, then the clump would be clearly detectable as an anomalous, bright gamma-ray source with the \textit{Fermi}-LAT. So it is very much possible to distinguish between dark matter origination of positrons and that from nearby GMCs considered in this work and/or pulsars considered in the literature.

The only other two candidates remaining for explaining the positron excess are pulsars considered in previous literature and GMCs considered in this work. Both of the models where pulsars are considered and model given in this work, where GMCs are considered, have successfully produced positron spectrum and positron fraction. Nearby pulsars and GMCs can both induce anisotropy and as seen from figure \ref{fig12}, anistropy from both of these models are below the \textit{Fermi}-LAT upper limits. Also, anisotropies calculated from both of these models supersede the anisotropy threshold for detection at 2$\sigma$ level. So, none of these models can be excluded in terms of anisotropy yet. However, with the development of better instrumentation, future observatories should be able to constrain these upper limits to a point where any anisotropy in the sky can be clearly discerned from isotropic background due to diffusion. Already \cite{fang18} has shown that a next generation CR observatory, high-energy cosmic-radiation detection (HERD) facility is expected to be better capable of detecting anisotropy than \textit{Fermi}-LAT. Since, from figure \ref{fig14}, it can be seen that nearby GMCs considered in this work and the pulsars are positioned at different RA and dec in the sky, it will be possible to distinguish anisotropy signals coming from those directions. So in future using updated, next generation instruments, based on anisotropies in different positions of the sky, it will be possible to unambiguously discern predictions from the model introduced in this work from that of the pulsar scenario.}

\section{Conclusion}
\label{sec:6}

The CR positron flux measured in  GeV$^2$  m$^{-2}$ sec$^{-1}$ sr$^{-1}$ rises with energy and peaks near 200 to 300 GeV. CR positrons are secondary particles produced in interactions of CR protons and heavy nuclei with hydrogen gas in ISM, and also in GMCs. It is difficult to explain the rise in CR positron flux unless there are sources close to the Earth. 
Earlier, pulsars and DM have been suggested as the origin of the rising positron flux or excess. {In fact recently, \cite{hooper18} has showed a complete solution in terms of Pulsar scenario.} In this work we discuss an alternative, self-consistent scenario of CR propagation, where CR positrons are produced in nearby GMCs in CR interactions and contribute significantly to the observed positron excess. CR proton and antiproton fluxes, B/C ratio, $^{10}$Be/$^{9}$Be ratio, electron, positron fluxes and positron fraction calculated using our model fit well to the observed data from different observations considered in this work. Thus we conclude that  nearby GMCs may play an important role in explaining the positron spectrum over the entire energy range of 1 to 1000 GeV.

\section*{References}

\bibliography{mybibfile}

\end{document}